\documentclass[12pt]{iopart}
\usepackage{graphicx}
\usepackage{hyperref}
\usepackage{bm}
\usepackage{color}

\usepackage{iopams}

\newcommand{\nn}{\nonumber}

\begin{document}

\title{Estimation of transport coefficients of dense hadronic and quark matter}

\author{Debashree Sen$^1$, Naosad Alam$^{1,2}$, and Sabyasachi Ghosh$^3$}

\address{$^1$Physics Group, Variable Energy Cyclotron Centre, 1/AF Bidhan Nagar, Kolkata 700064, India}
\address{$^2$Department of Nuclear and Atomic Physics, Tata Institute of
Fundamental Research, Homi Bhabha Road, Mumbai 400005, India}
\address{$^3$Indian Institute of Technology Bhilai, GEC Campus, Sejbahar, Raipur 492015, Chhattisgarh, India}
\ead{debashreesen88@gmail.com, naosadphy@gmail.com, sabyaphy@gmail.com}

\date{\today}

\begin{abstract}

We calculate the transport coefficients like shear viscosity and electrical conductivity with respect to density of dense hadronic and quark matter. By considering the simple massless limit for the quark matter and two different effective models for the hadronic matter, we have estimated the transport coefficients of the two phases separately. We have sketched the density profile of the transport coefficients in two parts viz. the phase-space part and the relaxation time part. By calculating the shear viscosity to density ratio, we have also explored the nearly perfect fluid domain of the quark and hadronic matter.

\end{abstract}

%
%
%
%

\section{Introduction}
\label{Intro}
 
At high density the composition and interaction of matter is still inconclusive from experimental perspectives. Therefore the understanding of the properties of dense matter is largely dependent on various theoretical models. Theoretically, hadronic matter may undergo phase transition to quark matter at high density or at high temperature. This hadron-quark transition density largely depends on the various theoretical models and the interactions considered. In the very low temperature and high baryon density domain of quantum chromodynamics (QCD) phase diagram, the first-principle predictions via lattice QCD (LQCD) calculations are missing due to the infamous sign problem~\cite{LQCD_sign}. However, the high temperature and low/vanishing baryon density domain of the QCD phase diagram is well studied for more than three decades~\cite{LQCD_88,LQCD_2010}. One can look at~\cite{LQCD_21} for its latest status. They conclude a cross-over type phase transition, which has been alternatively realized from the estimations of thermodynamic quantities at low and high temperature ranges with hadron resonance gas (HRG) model~\cite{HRG} and finite temperature perturbative QCD (pQCD) calculations~\cite{PQCD1,PQCD2}, respectively. Similar kind of mapping in the low/vanishing temperature and high baryon density domain may be possible~\cite{NS_QM1} by fusing ultra-high density (approximately 10 times larger than the hadronic saturation density) pQCD calculations~\cite{PQCD_mu1,PQCD_mu2,PQCD_mu3} and low density hadronic model calculations. In this context, the present work attempts to compare the estimations of transport coefficients, obtained from the standard massless outcome for the quark phase~\cite{Chodos} and the hadronic phase using two relativistic mean field (RMF) models viz. the effective chiral model~\cite{Sen2,Sen,Sen3,Sen5,Sen6,Sen4,Sahu2000,Sahu2004,TKJ} and the RMF model of ~\cite{Agrawal12,Sulaksono14,Alam15,Alam17}. At high density, the formation of heavier and strange baryons like the hyperons is theoretically possible. However, for simplicity and also due to the uncertainty in the hyperon couplings, we consider only the nucleons in the hadronic phase. Both the hadronic models adopted in this work, have been explored thoroughly to construct the equation of state (EoS) of dense matter and successfully determine the structural properties of neutron stars (NSs) or hybrid stars (HSs) in the light of recent constraints from various astrophysical observations \cite{Sen2,Sen,Sen3,Sen5,Sen6,Alam15,Alam17}. For the quark phase we consider the massless two flavor quark matter with u and d quarks. Similar to the hadronic phase we do not consider the strange degree of freedom (the s quark) in the quark phase. The present work is aimed to compare the estimations of the transport coefficients of both hadronic and quark matter in the high baryon density and low/vanishing temperature domain of QCD phase diagram. The motivation for this investigation comes from the equivalent pattern of LQCD thermodynamics~\cite{LQCD_2010,LQCD_21} and normalized transport coefficients~\cite{CAI_JD_SG} in the high temperature and low/vanishing baryon density domain of QCD phase diagram. If we analyze the temperature ($T$) profile of the thermodynamical quantities like pressure $P$, energy density $\epsilon$, entropy density $s$ and transport coefficients like shear viscosity $\eta$, electrical conductivity $\sigma$ for massless quark matter, then we can find the proportional relations $P=\frac{1}{3}\epsilon=\frac{1}{4}TS\propto T^4$ and $\eta\propto \tau T^4$, $\sigma\propto\tau T^2$, where $\tau$ is relaxation time of massless quark matter. So their normalized values $P/T^4$, $\epsilon/T^4$, $s/T^3$, $\eta/(\tau T^4)$ and $\sigma/(\tau T^2)$ will appear as horizontal line against $T$-axis and they can be marked as their upper or massless or the Stefan-Boltzmann (SB) limits. At very high $T$, these limiting values can be reached. As we go from high to low $T$, the values of thermodynamical quantities and transport coefficients will decrease and their maximum decrement will be noticed around quark-hadron transition temperature~\cite{CAI_JD_SG}. Here we are interested to find similar kind of graphs along baryon density axis.  

 Literature presents a long list of works~\cite{1950,1964,1970,1976,Baym1969,Yakovlev,Ewart,Raikh,Potekhin1999,Shternin2008, Banik2010,Banik2011,Schmitt2017,Shternin2020,McLaughlin,Shternin2013,Shternin2017, Shang2020,Nandi2018,Shternin2021} that have concentrated on the microscopic calculation of transport coefficients of the dense hadronic matter system. Refs.~\cite{1950,1964,1970,1976} are evident that the microscopic calculations of transport coefficients of dense hadronic matter is quite an old topic of research but it is still cultivated in recent time as reflected in~\cite{Schmitt2017,Shternin2020,McLaughlin,Shternin2013,Shternin2017,Shang2020, Nandi2018,Shternin2021}. For the calculation of transport coefficients of neutron star matter at high density, one also includes the electro-magnetic interaction part, whose time scale far away from QCD time scale. Many works~\cite{1950,1964,1970,1976,Baym1969,Yakovlev,Ewart,Raikh,Potekhin1999,Shternin2008, Banik2010,Banik2011,Schmitt2017,Shternin2020,McLaughlin,Shternin2013,Shternin2017, Shang2020,Nandi2018,Shternin2021} can be found in this regard, many of which have considered the contribution of the electrons and muons apart from the nucleons due to consideration of beta stable neutron star matter ~\cite{Shternin2008,Shternin_18}. However, in the present work we focus only on hadronic matter contribution and calculate the transport properties of dense hadronic matter. In the present work the transport coefficients estimations are obtained within QCD time scale (fm) as we have considered only the strong interaction related to dense hadronic and massless quark matter separately. The present work is focused within the QCD time scale for comparing the hadronic and quark phases pattern in terms of momentum and charge transportation due to QCD interactions. If one concentrates on the hadronic matter contribution of the works~\cite{Schmitt2017,Shternin2020,McLaughlin,Shternin2013,Shternin2017,Shang2020, Nandi2018,Shternin2021}, then it is noticed that the main ingredients for the calculation of the transport coefficients of hadronic matter are the effective masses of the nucleons and the quasiparticle relaxation time. Both quantities can have medium modification, which can be different in different models of many-body approach~\cite{Schmitt2017}. Owing to this model-dependent estimation scope, in the present work we have attempted to estimate the transport coefficients like shear viscosity and electrical conductivity of dense hadronic matter by using two different hadronic models. In this regard, it is just an alternative model estimation but we have followed an unique presentation of the transport coefficients along the density axis, which is generally adopted in the scenario of heavy-ion collisions~\cite{CAI_JD_SG}, where the normalized transport coefficients are studied with respect to temperature to understand the nearly perfect fluid nature of hadronic matter.  
 
  This article is organized as follows. Next in the Formalism section (\ref{sec:Form}), we have briefly addressed the two different models for the hadronic phase and the MIT Bag model for the massless quark phase in three different subsections (\ref{sec:Had1}), (\ref{sec:Had2}) and (\ref{sec:Bag}), respectively. Then in the Sec.~(\ref{sec:Tr}), the formalism of relaxation time approximation to calculate the transport coefficients is briefly addressed. After obtaining the final expressions of transport coefficients and the effective kinematic informations of the hadronic and the quark phases in the formalism part, we have shown their variations with respect to density in the result section (\ref{sec:Res}) along with detailed discussion. Finally in Sec~(\ref{sec:Sum}), we have summarized our findings.  


\section{Formalism of density-dependent hadronic and quark phases}
\label{sec:Form}

In this section, we will address briefly the two hadronic models calculations and the standard massless or zero-mass quark matter (z-MQM) calculations respectively. We have adopted two different hadronic models, whose brief details are given in the next two subsections (\ref{sec:Had1}), (\ref{sec:Had2}). In the third subsection (\ref{sec:Bag}), brief discussion of the z-MQM calculation for quark phase is given.

\subsection{Hadronic Phase: Hadronic model - 1}
\label{sec:Had1}
Considering only the nucleons as the baryonic degrees of freedom, the Lagrangian density for the effective chiral model \cite{Sahu2000,Sahu2004,TKJ} is given by
\begin{eqnarray}
\mathcal{L} = \overline{\psi} \Biggl[ \left(i \gamma_{\mu} \partial^{\mu} - g_{\omega}~ \gamma_{\mu} \omega^{\mu} -\frac{1}{2} g_{\rho}~ \overrightarrow{\rho_{\mu}} \cdot \overrightarrow{\tau} \gamma^{\mu} \right)-g_{\sigma} \left(\sigma + i \gamma_5 \overrightarrow{\tau} \cdot \overrightarrow{\pi} \right) \Biggr] \psi \nonumber \\
\hspace*{-1cm}+ \frac{1}{2} \left(\partial_{\mu} \overrightarrow{\pi} \cdot \partial^{\mu} \overrightarrow{\pi} + \partial_{\mu} \sigma ~ \partial^{\mu} \sigma \right) -{\frac{\lambda}{4}} \left(\chi^2-\chi_0^2\right)^2 - \frac{\lambda B}{6} (\chi^2-\chi_0^2)^3 - \frac{\lambda C}{8}(\chi^2-\chi_0^2)^4 \nonumber \\
- \frac{1}{4}F_{\mu\nu}F^{\mu\nu} 
+\frac{1}{2} {g_{\omega}}^2~\chi^2~\omega_\mu \omega^\mu 
- \frac{1}{4}~\overrightarrow{R_{\mu\nu}} \cdot 
\overrightarrow{R^{\mu\nu}}+\frac{1}{2}~m_\rho^2 ~\overrightarrow{\rho_\mu} \cdot \overrightarrow{\rho^\mu}~,
\protect\label{Lagrangian1}
\end{eqnarray}
where, $\psi$ is the nucleon isospin doublet. The nucleons interact with each other via the scalar $\sigma$ meson, the vector $\omega$ meson (783 MeV) and the isovector $\rho$ meson (770 MeV) with corresponding coupling strengths $g_{\sigma}$, $g_{\omega}$ and $g_{\rho}$, respectively. As mean field treatment is considered, the pions do not contribute. The model is based on chiral symmetry with the $\sigma$ and the pseudo-scalar $\pi$ mesons as chiral partners and $\chi^2 = (\pi^2 + \sigma^2 )$. The $\sigma$ field attains a vacuum expectation value (VEV) $\sigma_0=x_0$ with the spontaneous breaking of the chiral symmetry at ground state \cite{VolKo}. The masses of the nucleons ($m$) and the scalar and vector mesons can be expressed in terms of $\chi_0$ as 
\begin{eqnarray}
m = g_{\sigma} \chi_0,~~ m_{\sigma} = \sqrt{2\lambda}~ \chi_0,~~
m_{\omega} = g_{\omega} \chi_0~,
\end{eqnarray}
where, $\lambda=({m_\sigma}^2 -{m_\pi}^2)/(2{f_\pi}^2)$ is derived from chiral dynamics. The $f_\pi$, being the pion decay constant, relates to the VEV of $\sigma$ field as $<\sigma>=\sigma_0~=f_\pi$ \cite{Sahu2000,TKJ}. Since in mean field approximation, $<\pi> = 0$ and the pion mass becomes $m_{\pi} = 0$, the explicit contributions of the pions do not play any role in the interactions and also in the expression of $\lambda$. The term $\frac{1}{2} {g_{\omega}}^2 \chi^2 \omega_\mu \omega^\mu$ in Eq.~\ref{Lagrangian1} implies an explicit dependence of the nucleon effective mass on both the scalar and the vector fields and this is one of the salient features of the present model. The isospin triplet $\rho$ mesons are incorporated to account for the asymmetric hadronic matter. An explicit mass term for the isovector $\rho$ meson $\frac{1}{2}~m_\rho^2 ~\overrightarrow{\rho_\mu} \cdot \overrightarrow{\rho^\mu}$ is chosen following~\cite{Sahu1993,Sahu2000,Sahu2004,TKJ}. The coupling strength of the $\rho$ mesons with the nucleons is obtained by fixing the symmetry energy coefficient $J = 32$ MeV at hadronic saturation density $\rho_0$. In terms of the baryon density $\rho$ and the Fermi momentum $k_{F}=(6\pi^2 \rho/{\gamma})^{1/3}$ ($\gamma=2$ is degeneracy factor of nucleon), the isovector coupling strength is related to $J$ as
\begin{eqnarray}
J = \frac{C_{\rho}~ k_{F}^3}{12\pi^2} + \frac{k_{F}^2}{6\sqrt{(k_{F}^2 + m^{* 2})}}
\label{Crho}
\end{eqnarray}
where, $C_{\rho} = g^2_{\rho}/m^2_{\rho}$ and $m^*$ is the nucleon effective mass. The scalar density is obtained as
\begin{eqnarray}
\hspace*{-1cm}\rho_S=<\overline{\psi}\psi>=\frac{1}{ \pi^2} \Bigg[\int^{k_{Fn}}_0 dk_n ~k_n^2 \frac{m^*}{\sqrt{k_n^2 + {m^{*}}^2}}+\int^{k_{Fp}}_0 dk_p ~k_p^2 \frac{m^*}{\sqrt{k_p^2 + {m^{*}}^2}}\Bigg]
\end{eqnarray} 
while the baryon density for asymmetric hadronic matter is given as
\begin{eqnarray} 
\rho=<\psi^\dagger\psi>=\rho_n + \rho_p=\frac{1}{\pi^2} \Bigg[\int^{k_{Fn}}_0 dk_n ~k_n^2 + \int^{k_{Fp}}_0 dk_p ~k_p^2\Bigg].
\end{eqnarray}
 Symmetric nuclear matter (SNM) is defined as hadronic matter with equal number of neutrons and protons (N=Z) while at high density matter becomes asymmetric (N$>>$Z). The effective nucleon chemical potential is given as
\begin{eqnarray}
\mu_B=\sqrt{k_{F}^2 + {m^*}^2} + g_{\omega}\omega_0 + g_{\rho}I_{3B}\rho_{03}~,
\label{chempot1}
\end{eqnarray}
 where, $I_{3B}$ (with $B=n,p$) is the third component of isospin of the individual nucleons and $\omega_0$ and $\rho_{03}$ are the mean field values of the vector and isovector mesons, respectively. In Eq.~\ref{chempot1} the chemical potential of the individual nucleons in absence of the meson field terms is given by the first term $\mu_0=\sqrt{k_{F}^2 + {m^*}^2}$ which is modified into the effective chemical $\mu_B$ due to the interaction between the nucleons via exchange of the mesons as seen from the last two terms of Eq.~(\ref{chempot1}). So the effective chemical potential $\mu_B$ (where, $B=n,p$) of the individual nucleons differs by the third component of isospin of the individual nucleons $I_{3B}$ as seen from Eq.~(\ref{chempot1}).

The scalar EoM of the scalar field is

\begin{eqnarray} 
(1-Y^2)-\frac{B}{C_{\omega}}(1-Y^2)^2+\frac{C}{C_{\omega}^2}(1-Y^2)^3 + \frac{2~C_{\sigma}~C_{\omega}~\rho^2}{m^2 ~Y^4} - \frac{2~C_{\sigma}~\rho_{S}}{m~ Y}=0
\label{scalar_field}
\end{eqnarray} 
Here $C_i={g_i}^2/{m_i}^2$ are the scaled couplings ($i=\sigma~\&~\omega$, $m_i$ being the mass of the mesons) and $Y=m^*/m$. $B$ and $C$ are the coefficients of higher order scalar field terms. The EoM of the vector field is given as

\begin{eqnarray}
\omega_0=\frac{\rho}{g_{\omega} \chi^2}
\label{vector_field}
\end{eqnarray}
while that of the isovector field is

\begin{eqnarray} 
\rho_{03}=\sum_{B}\frac{g_{\rho}}{m_\rho^2}I_{3_B}\rho_B 
\label{isovector_field}
\end{eqnarray} 

The five model parameters $C_{\sigma}, C_{\omega}, C_{\rho}$ and the coefficients $B~ \&~ C$ of higher order scalar field terms are determined by reproducing the properties of SNM at saturation density $\rho_0$. The detailed procedure of obtaining these model parameters can be found in \cite{TKJ}. For the present work the parameter set is chosen from \cite{TKJ} and is presented in table \ref{table-HM1} below along with the SNM properties yielded by this parameter set.

\begin{table}[ht!]
\begin{center}
\caption{Model parameters chosen for the present work (adopted from \cite{TKJ}).}
\setlength{\tabcolsep}{15.0pt}
\begin{center}
\begin{tabular}{cccccccc}
\hline
\hline
\multicolumn{1}{c}{$C_{\sigma}$}&
\multicolumn{1}{c}{$C_{\omega}$} &
\multicolumn{1}{c}{$C_{\rho}$} &
\multicolumn{1}{c}{$B/m^2$} &
\multicolumn{1}{c}{$C/m^4$} &\\
\multicolumn{1}{c}{($\rm{fm^2}$)} &
\multicolumn{1}{c}{($\rm{fm^2}$)} &
\multicolumn{1}{c}{($\rm{fm^2}$)} &
\multicolumn{1}{c}{($\rm{fm^2}$)} &
\multicolumn{1}{c}{($\rm{fm^2}$)} & \\
\hline
6.772  &1.995  & 5.285 &-4.274   &0.292  \\
\hline
\hline
\multicolumn{1}{c}{$m^{*}/m$}&
\multicolumn{1}{c}{$K$} & 
\multicolumn{1}{c}{$B/A$} &
\multicolumn{1}{c}{$J$} &
\multicolumn{1}{c}{$L_0$} &
\multicolumn{1}{c}{$\rho_0$} \\
\multicolumn{1}{c}{} &
\multicolumn{1}{c}{(MeV)} &
\multicolumn{1}{c}{(MeV)} &
\multicolumn{1}{c}{(MeV)} &
\multicolumn{1}{c}{(MeV)} &
\multicolumn{1}{c}{($\rm{fm^{-3}}$)} \\
\hline
0.85  &303  &-16.3   &32  &87  &0.153 \\
\hline
\hline
\end{tabular}
\end{center}
\protect\label{table-HM1}
\end{center}
\end{table} 

Since the nucleon effective mass $m^{*}$ for this model depends on both the scalar and vector fields, it is quite high compared to well-known RMF models. Also at high density, unlike RMF models, the value of $m^{*}$ increases after a certain high value of density \cite{Sahu2004,TKJ,Sen2,Sen5}. This is due the dominance of vector potential at such density. Moreover, at high density the higher order terms of scalar field with coefficients $B$ and $C$ and the mass term of the vector field of the present model also become highly non-linear and dominant \cite{Sahu2004,TKJ,Sen2,Sen5}. The hadronic imcompressibility $K$ obtained with the chosen parameter set, though consistent with the results of \cite{Stone2} but it is larger than estimated in \cite{Khan2013,Garg}. The other SNM properties like the binding energy per nucleon $B/A$, the symmetry energy $J$ and the saturation density $\rho_0$ match well with the estimates of Refs.~\cite{Dutra2014,Stone}. The slope parameter $L_0$ is also quite consistent with the range specified by~\cite{Dutra2014,Fattoyev,Zhu2018}. The same model parameter set has also been adopted earlier in~\cite{Sen,Sen2,Sen3,Sen4,Sen5,Sen6,Sen7} to successfully investigate different properties of NSs as well as HSs in the light of various constraints specified on their structural properties.

\subsection{Hadronic Phase: Hadronic model - 2}
\label{sec:Had2}

In the conventional RMF theory \cite{Walecka74,Boguta77,Serot86,Furnstahl87,Muller96, Lalazissis97,Serot97,Agrawal12,Sulaksono14,Alam15,Alam17} nucleons are treated as elementary particles and interactions between the nucleons are mediated by 
 the exchange of $\sigma$, $\omega$ and $\rho$ mesons. The $\sigma$ mesons give rise to the strong attractive force, while the $\omega$ mesons cause the strong repulsive force between the nucleons. In addition to this, several self and cross interaction terms between the mesons are also considered to yield the saturation properties correctly. The Lagrangian density for the extended RMF model can be written as,
\begin{eqnarray}
\label{eq:lden}
{\cal L}= {\cal L_{NM}}+{\cal L_{\sigma}} + {\cal L_{\omega}} + {\cal
L_{\mathbf{\rho}}} +{\cal L_{\sigma\omega\mathbf{\rho}}}~, 
\end{eqnarray} 
where, the Lagrangian ${\cal L_{NM}}$ describing the interactions of the nucleons with mass $m$ through the mesons is, 

\begin{eqnarray}
\label{eq:lbm}
{\cal L_{NM}} &=& \sum_{B=n,p} \overline{\psi}_{i}[i\gamma^{\mu}\partial_{\mu}- (m-g_{\sigma} \sigma)
-(g_{\omega }\gamma^{\mu} \omega_{\mu}+\frac{1}{2}g_{\mathbf{\rho}}\gamma^{\mu}\tau .\mathbf{\rho}_{\mu})]\psi_{i}.~~~~~~
\end{eqnarray}
Here, the sum is taken over the neutrons and protons and $\tau$ are the isospin matrices. The Lagrangians for the $\sigma$, $\omega$, and $\rho$ mesons including their self interaction terms can be written as,

\begin{eqnarray}
\label{eq:NLnuclagNM}
{\cal L_{\sigma}} &=&
\frac{1}{2}(\partial_{\mu}\sigma\partial^{\mu}\sigma-m_{\sigma}^2\sigma^2)
-\frac{{\kappa_3}}{6m}
g_{\sigma}m_{\sigma}^2\sigma^3-\frac{{\kappa_4}}{24m^2}g_{\sigma}^2 m_{\sigma}^2\sigma^4,\nonumber\\
{\cal L_{\omega}} &=&
-\frac{1}{4}\omega_{\mu\nu}\omega^{\mu\nu}+\frac{1}{2}m_{\omega}^2\omega_{\mu}\omega^{\mu}
+\frac{1}{24}\zeta_0 g_{\omega}^{2}(\omega_{\mu}\omega^{\mu})^{2},\nonumber\\
{\cal L_{\mathbf{\rho}}}& =&
-\frac{1}{4}\mathbf{\rho}_{\mu\nu}\mathbf{\rho}^{\mu\nu}+\frac{1}{2}m_{\rho}^2\mathbf{\rho}_{\mu}\mathbf{\rho}^{\mu}
\end{eqnarray}
The $\omega^{\mu\nu}$, $\mathbf{\rho}^{\mu\nu}$ are field tensors
corresponding to the $\omega$ and $\rho$ mesons, and can be defined as
$\omega^{\mu\nu}=\partial^{\mu}\omega^{\nu}-\partial^{\nu}\omega^{\mu}$
and $\mathbf{\rho}^{\mu\nu}=\partial^{\mu}\mathbf{\rho}^{\nu}-
\partial^{\nu}\mathbf{\rho}^{\mu}$.  
Here, $m_{\sigma}$, $m_{\omega}$ and $m_{\rho}$ are the masses of 
$\sigma, \omega$, and $\mathbf{\rho}$ mesons, respectively.
The cross interactions of
$\sigma, \omega$, and $\mathbf{\rho}$ mesons are described by ${\cal
L_{\sigma\omega\rho}}$ which can be written as,

\begin{eqnarray}
\label{eq:NLnuclagMX}
{\cal L_{\sigma\omega\rho}} & =&
\frac{\eta_1}{2m}g_{\sigma}m_{\omega}^2\sigma\omega_{\mu}\omega^{\mu}+ 
\frac{\eta_2}{4m^2}g_{\sigma}^2 m_{\omega}^2\sigma^2\omega_{\mu}\omega^{\mu}
+\frac{\eta_{\rho}}{2m}g_{\sigma}m_{\rho }^{2}\sigma\rho_{\mu}\rho^{\mu}\nonumber \\
&&+\frac{\eta_{1\rho}}{4m^2}g_{\sigma}^2m_{\rho }^{2}\sigma^2\rho_{\mu}\rho^{\mu}
+\frac{\eta_{2\rho}}{4m^2}g_{\omega}^2m_{\rho
}^{2}\omega_{\mu}\omega^{\mu}\rho_{\mu}\rho^{\mu}~.
\end{eqnarray}

 The field equations derived from the above Lagrangian can be solved self-consistently by adopting mean-field approximation, i.e., the meson-field operators are replaced by their expectation values. The effective mass of the nucleon is 
 
\begin{eqnarray}
m^*=m-g_{\sigma} \sigma
\label{eq:em_hm2}
\end{eqnarray}
and the equilibrium densities are defined as $\rho = \rho_p + \rho_n$ and $\rho_3 = \rho_p -\rho_n$. 
 
 The values of the coupling constants are usually determined in such a way that they yield appropriate values for finite nuclei properties (e.g. binding energy, charge radii) and various quantities associated with the hadronic matter at the saturation density. The expression for the effective chemical potential of the individual nucleons in this model is also same as that in HM1 (Eq.~\ref{chempot1}) and is given as
 
\begin{eqnarray}
\mu_B= \sqrt{k_{F}^2 + {m^*}^2} + g_{\omega}\omega_0 + g_{\rho}I_{3B}\rho_{03}
\label{eq:mu_hm2}
\end{eqnarray}
with $I_{3B}$ being the isospin 3-component of nucleon. The fields involved in the expression of effective mass (Eq.~\ref{eq:em_hm2}) and chemical potential (Eq.~\ref{eq:mu_hm2}) of nucleons can be obtained by solving the following field-equations self-consistently, 

\begin{eqnarray}
\sigma_0 & = & 
-\frac{g_{\sigma}}{m_{\sigma}^2} \displaystyle{\sum_{i=n,p} 
   \frac{1}{\pi^2} \int_0^{k_{F_i}} dk\,k^2\, 
   \frac{m_i^{*}}{\sqrt{k^2+{m_i^*}^2}}}
 -\frac{1}{m_{\sigma}^2}\Bigl[
\frac{{\kappa_3}}{2m}g_{\sigma}m_{\sigma}^2\sigma_0^2\nonumber \\ 
&&+\frac{{\kappa_4}}{6m^2}g_{\sigma}^2 m_{\sigma}^2\sigma_0^3
-\frac{\eta_1}{2m}g_{\sigma}m_{\omega}^2\omega_0^2
-\frac{\eta_2}{2m^2}g_{\sigma}^2 m_{\omega}^2\sigma_0\omega_0^2\nonumber \\
&&-\frac{\eta_{\rho}}{2m}g_{\sigma}m_{\rho }^{2}\rho_{03}^2
-\frac{\eta_{1\rho}}{2m^2}g_{\sigma}^2m_{\rho }^{2}\sigma_0\rho_{03}^2 \Bigr]~, \\  
\omega_0 & = &
  \frac{g_{\omega}}{m_{\omega}^2} \rho
 -\frac{1}{m_{\omega}^2}\Bigl[
\frac{1}{6}\zeta_0 g_{\omega}^{2}\omega_0^3
- \frac{\eta_1}{m}g_{\sigma}m_{\omega}^2\sigma_0\omega_0 \nonumber \\  
&& -\frac{\eta_2}{2m^2}g_{\sigma}^2 m_{\omega}^2\sigma_0^2\omega_0
-\frac{\eta_{2\rho}}{2m^2}g_{\omega}^2m_{\rho
}^{2}\rho_{03}^2\omega_0\Bigr]
,  \\                                                                     
\rho_{03}   &=&
   \frac{g_{\rho}}{m_{\rho}^2} 
   \rho_3
+\frac{1}{m_{\rho}^2}\Bigl[ \frac{\eta_{\rho}}{m}g_{\sigma}m_{\rho }^{2}\sigma_0\rho_{03}\nonumber
 +\frac{\eta_{1\rho}}{2m^2}g_{\sigma}^2m_{\rho }^{2}\sigma_0^2\rho_{03} \\
 &&+\frac{\eta_{2\rho}}{2m^2}g_{\omega}^2m_{\rho
}^{2}\omega_0^2\rho_{03}  \Bigr],\\                                                    
\end{eqnarray} 

We use the BSP parametrization \cite{Agrawal12} for our calculation, which describe the properties of finite nuclei very well. This parameterization includes the quartic order cross-coupling between $\omega$ and $\sigma$ mesons to model the high density behavior of the EoS.

\begin{table*}[ht!]
\centering
\caption{\label{tab:par_sets}
BSP parameter sets for the extended RMF model with the nucleon mass $m =$ 939.2 MeV.  }
\scalebox{0.8}{
\begin{tabular}{cccccccccccccc}
\hline
\hline
$g_\sigma/4\pi $&  $g_\omega/4\pi $& $g_\rho/4\pi $&  $\kappa_3 $& $\kappa_4$&  $\eta_1$&  $\eta_2$& 
$\eta_{\rho}$&$\eta_{1\rho}$&$\eta_{2\rho}$&$\zeta_0$&$m_\sigma/m$&$m_\omega/m$&$m_\rho/m$\\
\hline
0.8764& 1.1481 &1.0508&1.0681&14.9857&0.0872&  3.1265&0.0&0.0&53.7642&0.0&0.5383&0.8333&0.8200\\
\hline
\hline
\end{tabular}
}
\end{table*}

\begin{table}[ht!]
\centering
\caption{\label{tab:nm_pro}
Properties of the hadronic matter at the saturation density.}
\begin{tabular}{cccccc}
\hline
\hline
$m^*/m$ & $B/A$&$\rho_0$& $K$& $J$&$ L_0$\\
  & (MeV)& (fm$^{-3}$)& (MeV)& (MeV)&(MeV)\\
\hline
   0.60 & -15.9&  0.149& 230&   28.83 &  50  \\
\hline
\hline
\end{tabular}
\end{table}

 In the present work we consider asymmetric matter motivated by the asymmetry of dense neutron star matter. In typical dense neutron star matter the proton content is $\sim$(10-15)\% of the total matter. Therefore in the present work we consider the proton fraction to be 15\% in case of both the hadronic models HM1 and HM2. However, as discussed in the Introduction section \ref{Intro}, we do not consider the leptonic contribution in the hadronic sector of the present work to calculate the transport coefficients.

\subsection{Quark Phase}
\label{sec:Bag}

We consider the MIT bag model~\cite{Chodos,Glendenning} with u and d quarks to describe the pure massless quark phase. 
The chemical potential of the quarks is respectively given as
\begin{eqnarray}
\mu_f=\sqrt{{k_{F}}_f^2 + m_f^2}
\label{chempot_q}
\end{eqnarray}
where, $k_f$ is the Fermi momenta of individual flavors $f=$ u and d. $m_f$ is the mass of individual quarks. We consider $m_u=$ 2.2 MeV and $m_u=$  4.7 MeV, which are low enough to consider the quark matter to be massless in absence of the massive s quark. The total baryonic density is given as
\begin{eqnarray}
\rho=\frac{\rho_u+\rho_d}{3}=\frac{1}{3}(2\times 3) \frac{1}{6\pi^2} \Bigg[{k_{Fu}^3} + {k_{Fd}^3}\Bigg],
\protect\label{density_upq}
\end{eqnarray}
where spin, color degeneracy factors of quark matter are considered as 2, 3 respectively and $k_{Fu}$, $k_{Fd}$ are Fermi-momentum of the u and d quarks respectively. The factor 1/3 comes to conserve the net baryon number since the MIT Bag model is based on deconfinement of hadrons (nucleons) and the fact that there are three quarks per nucleon, each having baryon number 1/3.

\section{Framework of transport coefficients for hadrinic and quark phases}
\label{sec:Tr}

Here we will address the standard kinetic theory framework for obtaining the transport coefficients like shear viscosity $\eta$ and electrical conductivity $\sigma$, which are basically the part of the dissipation component of any many body system or medium or fluid. Let us first derive the expression of shear viscosity. Considering the matter as a dissipative fluid, one can express energy-momentum tensor in macroscopic form as
\begin{eqnarray}
T^{\mu \nu}={T_0}^{\mu\nu} + {T_D}^{\mu\nu}~,
\end{eqnarray}
where, ideal (${T_0}^{\mu\nu}$) and dissipation (${T_D}^{\mu\nu}$) parts in terms of fluid quantities like fluid four velocity $u^\mu$, energy density $\epsilon$, pressure $P$ and metric tensor $g^{\mu\nu}$ can be expressed as 
\begin{eqnarray}
{T_0}^{\mu \nu}=-g^{\mu\nu} + (\epsilon +P) u^{\mu}u^{\nu} 
\end{eqnarray}
and
\begin{eqnarray}
T_D^{\mu\nu}=\pi^{\mu\nu} +...=\eta^{\mu\nu\alpha\beta}~ {\mathcal{U}}_{\alpha \beta}^{\eta} +..
\label{TD_U}
\end{eqnarray}
respectively. In the expression of the dissipative part of the energy-momentum tensor, the viscous stress tensor is $\pi^{\mu\nu}$, velocity gradient tensor is ${\mathcal{U}}^{\mu \nu}_{\eta}= D^{\mu}u^{\nu} + D^{\nu}u^{\mu} -\frac{2}{3} \Delta^{\mu \nu} \partial_{\rho} u^{\rho}$, with 
$D^\mu= \partial^\mu - u^{\mu}u^{\sigma} \partial_{\sigma}$, $\Delta^{\mu\nu}=g^{\mu\nu}-u^\mu u^\nu$ and $\eta^{\mu\nu\alpha\beta}$ is shear viscosity tensor. Also, the dissipative part can have components of shear, bulk and thermal dissipation but here we are dealing only with the shear part. The notation of $(+... )$ indicates the other dissipation components. The picture of relativistic shear viscosity $\eta$ as a proportional constant between viscous stress tensor $\pi^{\mu\nu}$ and tangential fluid velocity gradient ${\mathcal{U}}^{\mu \nu}_{\eta}$ can be compared with Newton-Stoke law, applicable in the non-relativistic domain, where the shear viscosity is defined as the proportionality constant between shear stress and velocity gradient. 

 Considering the hadronic matter to be composed of nucleons, the microscopic expression of its energy-momentum tensor can be written in kinetic theory framework as
\begin{eqnarray}
T^{\mu \nu}= \gamma \int \frac{d^3 \vec{k}}{(2\pi)^3} \frac{k^{\mu}k^{\nu}}{E} (f_N + f_{\bar{N}})
\label{Tmunu}
\end{eqnarray}
where, $f_N$ and $f_{\bar{N}}$ are assumed as non-equilibrium distribution functions for nucleons $N$ and anti-nucleons ${\bar{N}}$. Now splitting $f_{N,{\bar{N}}}$ as the sum of equilibrium distribution $f_{N,{\bar{N}}}^0$ and a small deviation $\delta f_{N,{\bar{N}}}$ i.e, $f_{N,{\bar{N}}}= f_{N,{\bar{N}}}^0 + \delta f_{N,{\bar{N}}}$, one can separately identify the microscopic expressions of ${T_0}^{\mu\nu}$ and ${T_D}^{\mu\nu}$ parts in terms of the quantities like particle's four momenta $k^\mu$, degeneracy factor $\gamma$ etc. as
\begin{eqnarray}
{T_0}^{\mu \nu}=\gamma \int \frac{d^3 \vec{k}}{(2\pi)^3} \frac{k^{\mu}k^{\nu}}{E} (f^0_N + f^0_{\bar{N}})
\end{eqnarray}
and
\begin{eqnarray}
T_D^{\mu\nu}=\pi^{\mu\nu} +...=\gamma \int \frac{d^3 \vec{k}}{(2\pi)^3} \frac{k^{\mu}k^{\nu}}{E} (\delta f_N + \delta f_{\bar{N}}) +..
\label{TD_df}
\end{eqnarray}
Now, with the help of the standard relaxation time approximation (RTA) methods of Boltzmann transport equation, one can get~\cite{CAI_JD_SG}

\begin{eqnarray}
\delta f_{N,{\bar{N}}}=\frac{k^\alpha k^\beta}{E}\tau_{N,{\bar{N}}}\beta f^0_{N,{\bar{N}}}(1-f^0_{N,{\bar{N}}}){\mathcal{U}}_{\alpha \beta}^{\eta}~,
\label{df_U}
\end{eqnarray}
where, $\tau_{N,{\bar{N}}}$ are relaxation times and $E$ are energy of nucleon and anti-nucleon. Now, using Eq.~(\ref{df_U}) in Eq.~(\ref{TD_df}) and then comparing with Eq.~(\ref{TD_U}), we get the shear viscosity tensor as

\begin{eqnarray} 
\eta^{\mu\nu\alpha\beta}=\gamma \int \frac{d^3 \vec{k}}{(2\pi)^3} \frac{k^{\mu}k^{\nu}k^\alpha k^\beta}{E^2} \tau_{N,{\bar{N}}} [f^0_{N}(1-f^0_{N}) + f^0_{\bar{N}}(1-f^0_{\bar{N}})]/T
\end{eqnarray} 
Using tensor identity~\cite{CAI_JD_SG,Kapusta,Gavin}, one can obtain the isotropic expression as

\begin{eqnarray}
\eta=\frac{\gamma}{15} \int \frac{d^3 \vec{k}}{(2\pi)^3} \frac{k^4}{E^2} \tau_{N,{\bar{N}}} [f^0_{N}(1-f^0_{N}) + f^0_{\bar{N}}(1-f^0_{\bar{N}})]/T
\label{eta}
\end{eqnarray}
 If we analyze Eq.~(\ref{eta}), then we can find two components. One is relaxation time $\tau_{N,{\bar{N}}}$, whose order of magnitude will fix the strength of viscosity for the system. Another is the remaining part of the expression of $\eta$, which can be called as phase-space part of $\eta$ and it will basically decide the temperature and density dependence of $\eta$. The expression for any transport coefficient will have these two component structures.

Now, if we take our calculation from finite $T$ to the $T=0$ case, then the equilibrium Fermi-Dirac distribution takes the form of a step function as
\begin{eqnarray}
f_{N}^0 &=& 1~ {\rm{if}} E<\mu 
\nonumber\\
                  &=& 0~ {\rm{if}} E>\mu 
\nonumber\\
f_{\bar{N}}^0 &=& 1~ {\rm{if}} E<-\mu 
\nonumber\\
                  &=& 0~ {\rm{if}} E>-\mu   ~.               
\end{eqnarray}
The above conditions imply that the anti-nucleons do not contribute anymore to the positive energy for the $T=0$ case. We have to use the replacement

\begin{eqnarray}
\frac{\partial f^0_N}{\partial E} = - f^0_N(1-f^0_N)/T
\rightarrow 
\frac{\partial}{\partial E}\theta(\mu-E)=-\delta(E-\mu)~.
\end{eqnarray}
Using these replacements in Eq.~(\ref{eta}), we get

\begin{eqnarray} 
\eta &=& \frac{\gamma}{15} \int \frac{d^3 \vec{k}}{(2\pi)^3} \frac{k^4}{E^2} \tau_{N}\delta(E-\mu)
\nonumber\\
&=&\frac{\gamma}{30\pi^2}\tau_N\frac{(\mu^2-m^{*2})^{5/2}}{\mu}
\label{eta_mu}
\end{eqnarray} 
Here, $\mu$ is the weighted average of the effective chemical potentials of neutron and proton as obtained from Eq.~(\ref{chempot1}) and Eq.~(\ref{eq:mu_hm2}) for HM1 and HM2, respectively.

 Let us next come to another transport coefficient viz. the electrical conductivity, which is basically proportionality constant between electric current density and the field. This macroscopic definition comes from Ohm's law, which can be expressed in three dimensional notation as

\begin{eqnarray}
J^{i}= \sigma^{ij} E_{j}~,
\label{J_E}
\end{eqnarray}
where, the electrical conductivity tensor $\sigma^{ij}$ connect the electric field $E_j$ and the electric current density $J^i$. Realizing current due to electric field as a dissipation phenomenon, we can microscopically express it as 
\begin{eqnarray}
J^i= q\gamma \int \frac{d^3 \vec{k}}{(2\pi)^3} \frac{k^{i}}{E} \delta f_N~,
\label{J_df}
\end{eqnarray}
where, $q$ is the electric charge of nucleon, i.e. $q=0, e$ for neutron and proton. Here, the RTA methods of Boltzmann transport equation will again help us to give the form of $\delta f_N$~\cite{CAI_JD_SG}:

\begin{eqnarray} 
\delta f_{N}=\frac{k^j}{E}\tau_{N}\beta f^0_{N}(1-f^0_{N})qE_j~,
\label{df_E}
\end{eqnarray}
Using Eq.~(\ref{df_E}) in Eq.~(\ref{J_df}) and then comparing with Eq.~(\ref{J_E}), we get the conductivity tensor as

\begin{eqnarray} 
\sigma^{ij}= q^2\gamma \int \frac{d^3 \vec{k}}{(2\pi)^3} \frac{k^{i}k^j}{E^2}
\tau_{N}\beta f^0_{N}(1-f^0_{N})~,
\end{eqnarray} 
whose isotropic expression at $T=0$ will be
\begin{eqnarray} 
\sigma &=& \frac{q^2\gamma}{3} \int \frac{d^3 \vec{k}}{(2\pi)^3} \frac{k^2}{E^2} \tau_{N}\delta(E-\mu)
\nonumber\\
&=&\frac{q^2\gamma}{6\pi^2}\tau_N\frac{(\mu^2-m^{*2})^{3/2}}{\mu}
\label{sig_mu}
\end{eqnarray} 
The final expressions of shear viscosity and electrical conductivity for hadronic matter are given by Eqs.~(\ref{eta_mu}) and (\ref{sig_mu}), respectively. The neutrons will not contribute to the electrical conductivity given in Eq.~(\ref{sig_mu}). Hence only protons contribute to  the electrical conductivity and thus in Eq.~(\ref{sig_mu}) we have $\mu=\mu_p$ as the effective proton chemical potential, calculated from Eqs.~(\ref{chempot1}) and (\ref{eq:mu_hm2}) for HM1 and HM2, respectively. Moreover, in Eq.~(\ref{sig_mu}) only protons will contribute to the relaxation time i.e., $\tau_N=\tau_p$.

 When we apply these two expressions for quark phase, then nucleon effective mass $m^*$ will be replaced by quark mass and the degeneracy factor $\gamma$ will also be replaced by the corresponding quark degeneracy factor $g$ (say). Considering two flavor quark matter with u and d quarks, $g=3\times 2\times 2=12$ for Eq.~(\ref{eta_mu}) and $gq^2=3\times 2 \Big(\frac{4e^2}{9}+\frac{e^2}{9} \Big)=\frac{10e^2}{3}$ for Eq.~(\ref{sig_mu}). The values of $\eta$ and $\sigma$ for quark matter will be very close to their massless limits

\begin{eqnarray}
\eta &=&\frac{g}{30\pi^2}\tau_Q \mu^4
\nonumber\\
\sigma &=&\frac{g q^2}{6\pi^2}\tau_Q \mu^2,
\label{es_QM}
\end{eqnarray} 
whose, normalized values become constant. These massless limits may act as reference line at $T=0$ and finite $\mu$ case. 

 The relaxation time can also be calculated microscopically. For the degenerate scenario, the medium constituents will occupy all the energy levels from ${m^*}$ to $\mu$ and ideally they have zero probability to move outside $\mu$. However, in reality, due to very low $T$ instead of exactly $T=0$, we can have a small deviation from step function type distribution function. So medium constituents, having energy nearly Fermi energy $\mu$ and velocity nearly Fermi velocity $v_F=\sqrt{\mu^2 - {m^*}^2}/\mu$ will participate in the momentum transfer scattering process. So we can define relaxation time as 
 
\begin{eqnarray}
\tau_c &=& 1/[\sigma_s v_F \rho]
\nn\\
&=& \mu/[\sigma_s \sqrt{\mu^2-{m^*}^2} ~\rho]~,
\label{tau}
\end{eqnarray}
where, $\rho$ is density of the medium and $\sigma_s=4\pi a^2$ is cross section with scattering length $a$. Now depending upon the system, we have to use the inputs for the medium constituents to calculate their relaxation time. We first normalize the quantities $\eta$ and $\sigma$ as $\eta/(\tau_c\mu^4)$ and $\sigma/(\tau_c\mu^2)$ to obtain dimensionless forms. For electrical conductivity we consider $\tau_c=\tau_{cp}$ and $\mu=\mu_p$ for the scaling.
 

\section{Result and Discussions}
\label{sec:Res}

 We obtain the transport coefficients of the hadronic and quark phases based on the formalism discussed in the earlier section and study the variation of these coefficients with respect to density. The phenomenon of phase transition at vanishing temperature and the transition density is still inconclusive in literature. So rather than achieving phase transition, in the present work we compare the order of magnitudes of the transport coefficients for the hadronic and quark phases. 
\begin{figure}[!ht]
\centering
{\includegraphics[width=0.49\textwidth]{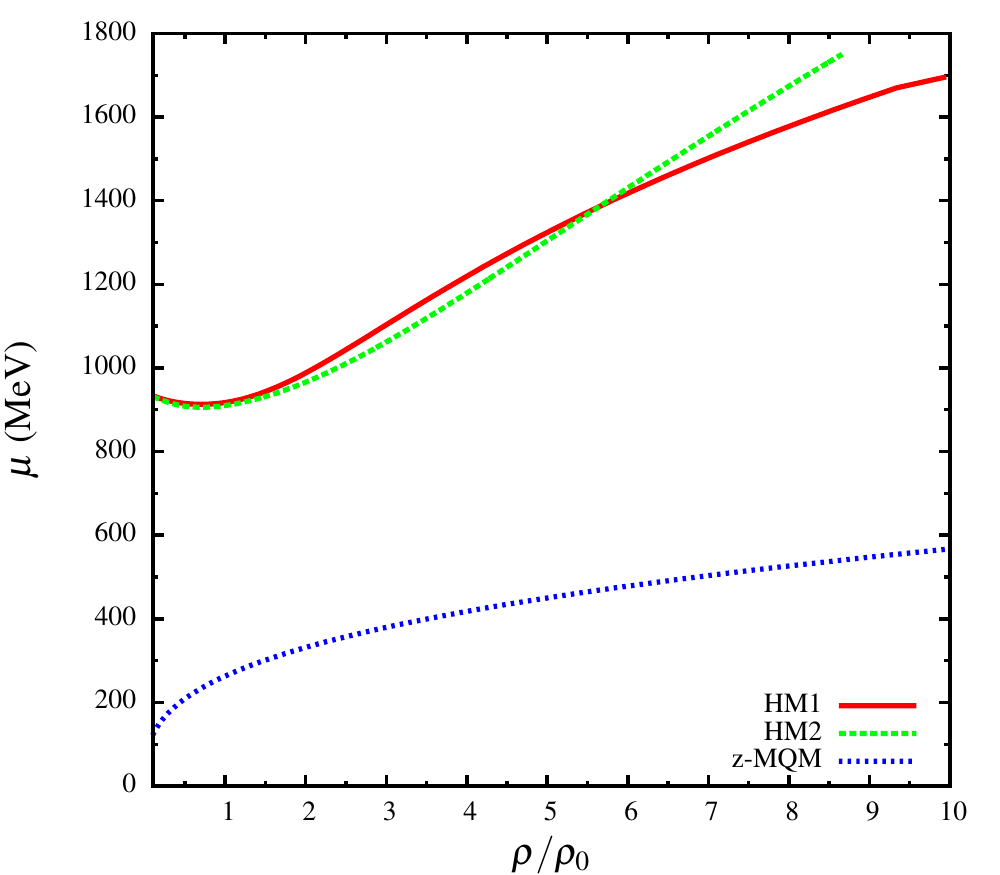}}
{\includegraphics[width=0.49\textwidth]{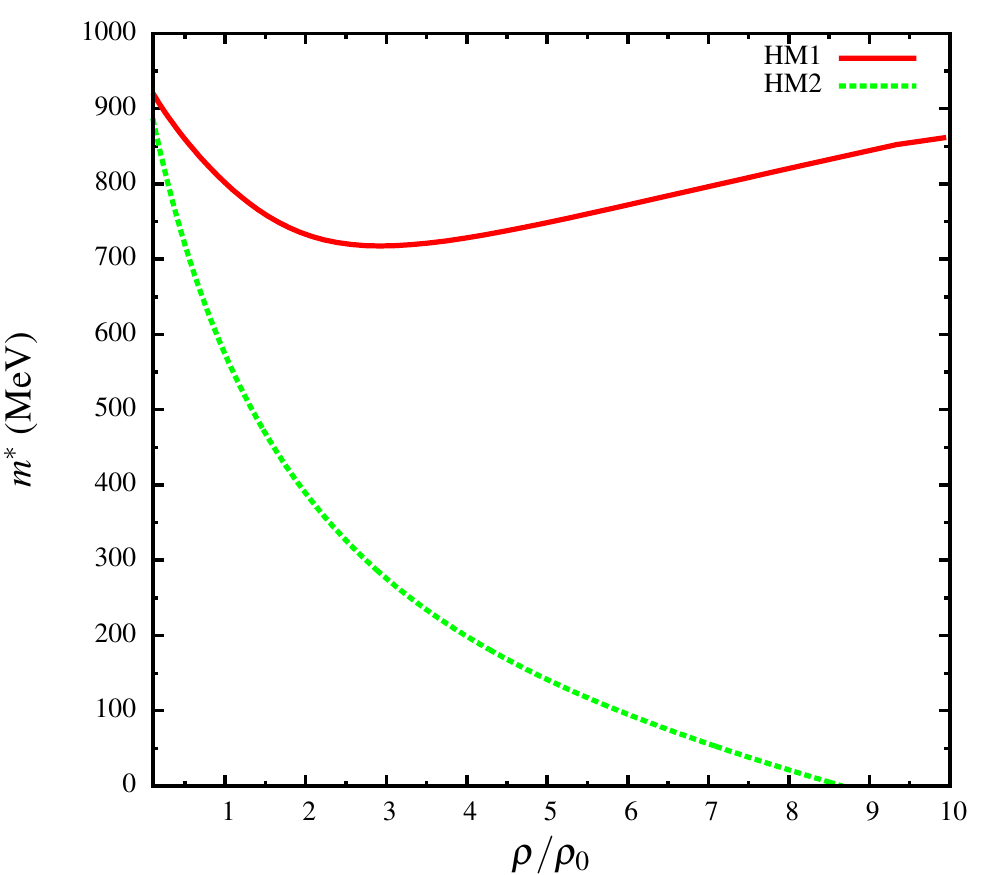}}
\caption{Weighted average of the effective chemical potential of hadronic matter with two different hadronic models HM1 and HM2 and zero-mass quark matter (z-MQM (left) and the effective nucleon mass of hadronic matter with HM1 and HM2 (right) vs normalized baryon density.}
\label{mu_mstr}
\end{figure}

\begin{figure}  
\centering
{\includegraphics[width=0.49\textwidth]{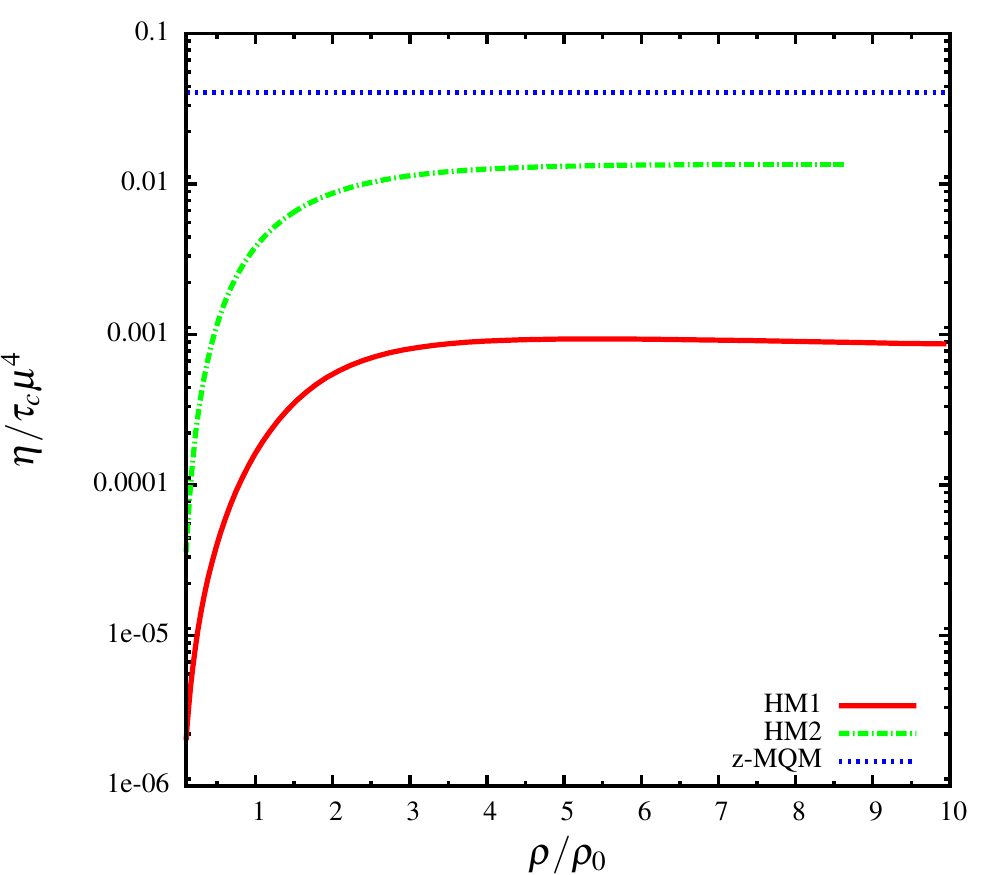}}
{\includegraphics[width=0.49\textwidth]{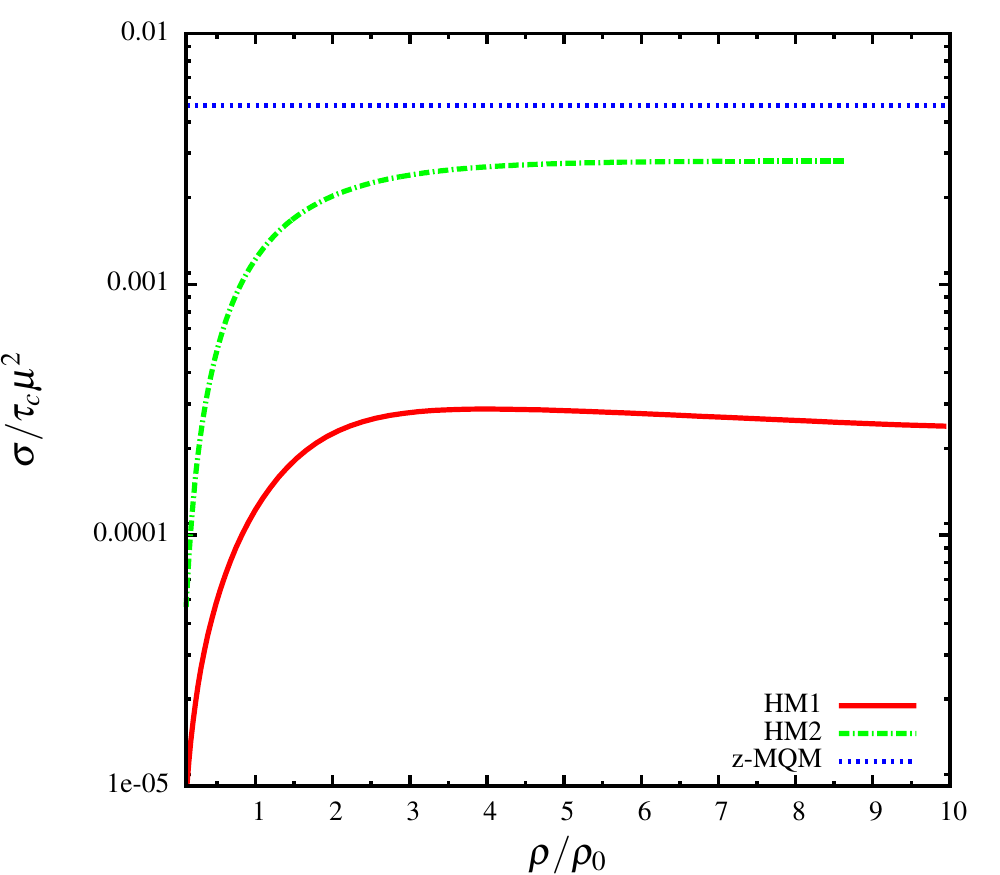}}
\caption{Normalized shear viscosity (left) and electrical conductivity (right) with respect to scaled baryon density of hadronic matter and zero-mass quark matter (z-MQM).}
\label{es_QH}
\end{figure}

 We calculate the shear viscosity of hadronic and quark matter using Eq.~(\ref{eta_mu}). The electrical conductivity for hadronic matter is calculated using Eq.(\ref{sig_mu}) while that of quark matter is calculated using Eq.~(\ref{es_QM}). We then normalize these quantities as $\eta/(\tau_c\mu^4)$ and $\sigma/(\tau_c\mu^2)$ to obtain dimensionless forms, where $\tau_c$ is the relaxation time of quark or hadronic matter in general i.e. $\tau_c=\tau_N$ for hadronic matter and $\tau_c=\tau_Q$ for quark matter. These quantities $\eta/(\tau_c\mu^4)$ and $\sigma/(\tau_c\mu^2)$ are plotted against the normalized density in left and right panels of Fig.~(\ref{es_QH}), respectively. We notice roughly 10-20 times difference between the hadronic model-1 (HM1) and hadronic model-2 (HM2) in terms of the estimations of $\eta/(\tau_c\mu^4)$ and $\sigma/(\tau_c\mu^2)$. This is because of the difference in chemical potential ($\mu$) and also the nucleon effective mass ($m^*$) between the two hadronic models. Both chemical potential and specially the effective mass are quite different for HM1 and HM2 as seen from Fig.~(\ref{mu_mstr}). Re-looking at Eqs.~(\ref{eta_mu}) and (\ref{sig_mu}), one can understand the numerical and physical impact of Fig.~(\ref{mu_mstr}) on Fig.~(\ref{es_QH}). The difference in the order of magnitudes between the two hadronic models reflects a numerical band of transport coefficients for hadronic matter along density axis. For both $\eta/(\tau_c\mu^4)$ and $\sigma/(\tau_c\mu^2)$, we notice jump in the estimated values of the hadronic models and the quark model (z-MQM). For shear viscosity, this jump is from 0.001 (HM1) or 0.02 (HM2) to 0.05 (z-MQM) and for electrical conductivity this jump is from 0.0003 (HM1) or 0.003 (HM2) to 0.005 (z-MQM). One may notice the saturation trend of this normalized transport coefficients, which zoom in basically its thermodynamical phase-space part (or in other words, the thermodynamical probability of shear and electrical charge transportation). The saturation profile of HM1 (red solid line) and HM2 (green dash line) have a mild increasing trend with $\mu$ due to their dependence $\eta \propto \tau_c\mu^4 (1-m^{*2}/\mu^2)^{5/2}$ and $\sigma\propto\tau_c\mu^2(1-m^{*2}/\mu^2)^{5/2}$. They saturate at different values due to the difference in $m^*$ between the two hadronic models.

On the other hand, the quark matter estimation follow the massless transport relation along the entire density axis $\eta \propto \tau_c\mu^4$ and $\sigma\propto\tau_c\mu^2$ since we have put exact massless limit for the quark phase. In actual case one should sketch the estimations of HM1 or HM2 at low density domain and the z-MQM estimation at high density domain and at some intermediate density quark-hadron phase transition will take place, connecting the estimations of the low and high density domains. However, due to lack of proper knowledge regarding phase transition density (which is highly model dependent), we restrict our focus only on the order of magnitude of the transport coefficients for two phases.  

 These estimations of the transport coefficients viz. the shear viscosity and the electrical conductivity at $T=0$ and $\mu\neq 0$ domain can be compared with the estimations of the transport coefficients at $T\neq 0$ and $\mu= 0$ domain, given in Ref.~\cite{CAI_JD_SG} and the references therein. The quark-hadron transition at $T\neq0$ and $\mu=0$ domain is understood by estimating the thermodynamics of massless quark gluon plasma (QGP) and hot pion gas for high and low temperature zones, respectively. However, lattice quantum chromodynamics (LQCD) calculations~\cite{LQCD_88,LQCD_2010,LQCD_21} went deeper and unfolded the crossover type nature of quark-hadron transition at $T\neq 0$ and $\mu= 0$ domain. The values of normalized thermodynamical quantities like $P/T^4$, $\epsilon/T^4$ etc. for massless QGP will act as reference line or upper limits of QCD matter for the $\mu=0$ case. The values of thermodynamical quantities, obtained from LQCD, remain quite lower than the massless QGP limit in low (or hadronic matter) temperature domain, which can be realized as non-perturbative aspects of QCD. Near transition temperature, their values increase to reach towards the massless limits but in a smooth crossover way instead of a first order phase transition kind of jump. Beyond the transition temperature, LQCD thermodynamical values still remain little suppressed with respect to their massless limits. This small suppression is also realized from the direction of finite temperature perturbative quantum chromodynamics (pQCD) theory, whose latest status can be found in~\cite{HTL_3L}. Similar investigation of thermodynamical variations with respect to density or chemical potential at $T=0$ are attempted over a very long time for understanding the properties of dense matter~\cite{Hatsuda}. Due to problems in LQCD calculations at finite $\mu$, the present knowledge of quark-hadron phase transition along $\mu$-axis is not quite converging like the understanding of phase transition along $T$-axis. In this context, in the present work we consider the effective hadronic models and the Bag model to get the order of magnitudes for different transport coefficients of the hadronic and quark phases along $\rho$ or $\mu$-axis. The normalized transport coefficients $\eta/(\tau_c\mu^4)$ and $\sigma/(\tau_c\mu^2)$ against $\rho$ or $\mu$-axis will exhibit thermodynamical phase-space of transportation, which follow similar suppression in hadronic $\mu$-domain (at $T=0$) as noticed for hadronic $T$-domain (at $\mu=0$) in~\cite{CAI_JD_SG}. In~\cite{CAI_JD_SG}, the normalized coefficients are chosen as $\eta/(\tau_c T^4)$ and $\sigma/(\tau_c T^2)$, while in the present work they are chosen as $\eta/(\tau_c\mu^4)$ and $\sigma/(\tau_c\mu^2)$. In both cases, around $10^{1-2}$ suppression is observed in hadronic $T$ or $\mu$ domain with respect to their massless limits. The suppressed values in both the cases can be realized as the non-pQCD effect in the phase-space of transportation. This equivalence between normalized phase-space of transport coefficients along $T$-axis and $\rho$-axis may be considered as an unique finding of present investigation.

 Up to now, the transport coefficients are normalized by relaxation time. So we basically see the phase space part of transport coefficients. However, the actual estimation of relaxation time will be important to know the absolute values of the transport coefficients. We now calculate the relaxation time for the hadronic and massless quark matter using Eq.~(\ref{tau}). We consider the effective nucleon mass and two possible orders of cross-section $\sigma_s=4\pi a^2\approx 5340~\rm{fm^2}$ and $ 12.56~\rm{fm^2}$ for hadronic matter case with scattering lengths $a\approx 20$ fm and $1$ fm. The former value of $\sigma_s\approx 5340~\rm{fm^2}$ or $a\approx 20$ fm is the isospin averaged cross section/scattering length of NN interactions, taken from~\cite{NN1,NN2}. This value is used for the calculation of relaxation time of nucleons at finite temperature in~\cite{SB_SG2}. Here, we have considered finite density system. We also consider another small value of scattering length $a=1$ fm.

\begin{figure}[!ht]
\centering
{\includegraphics[scale=1.0]{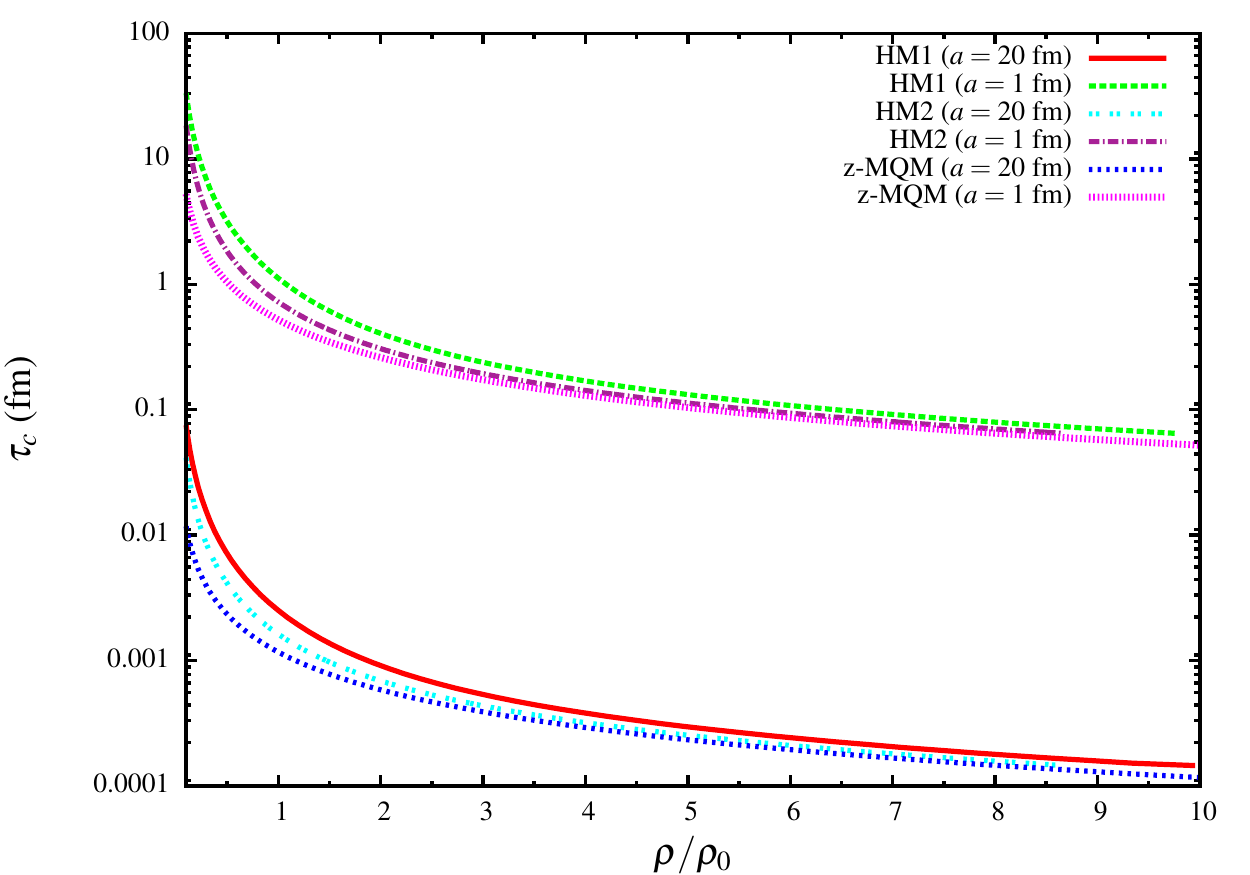}}
\caption{Relaxation time with respect to scaled baryon density of hadronic matter and zero-mass quark matter (z-MQM) for different values of cross-section.}
\label{tau_mu}
\end{figure}

 In Fig.~(\ref{tau_mu}), we have plotted the variation of $\tau_c$ for HM1 and HM2 and z-MQM. For hadronic matter we consider two different values of scattering lengths - one is guided by the vacuum scattering interaction strength and another is a guess value by assuming that scattering length may decrease with density. As a rough order of magnitude, we get $\tau_c\approx 10-0.1$ fm and $\tau_c\approx 0.01-0.0002$ fm for scattering lengths $a\approx 1$ fm and $a=20$ fm, respectively. One should notice that in this context the ranking of $\tau_c$ is HM1 $>$ HM2 $>$ z-MQM. This is because of the inverse relation of relaxation time with thermodynamic quantity like density $\rho$ as seen from Eq. \ref{tau}.

\begin{figure}
\centering
{\includegraphics[width=0.49\textwidth]{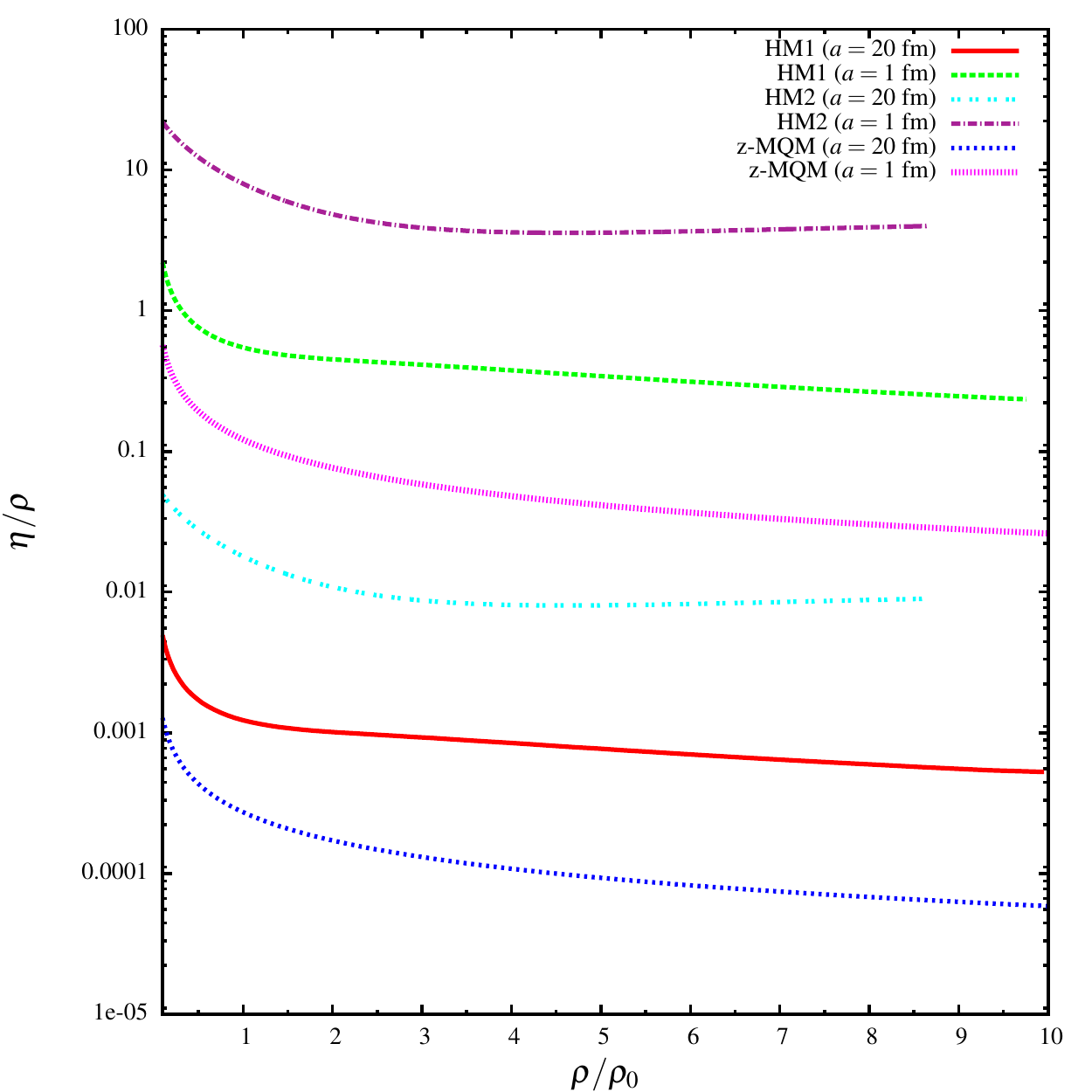}}
{\includegraphics[width=0.49\textwidth]{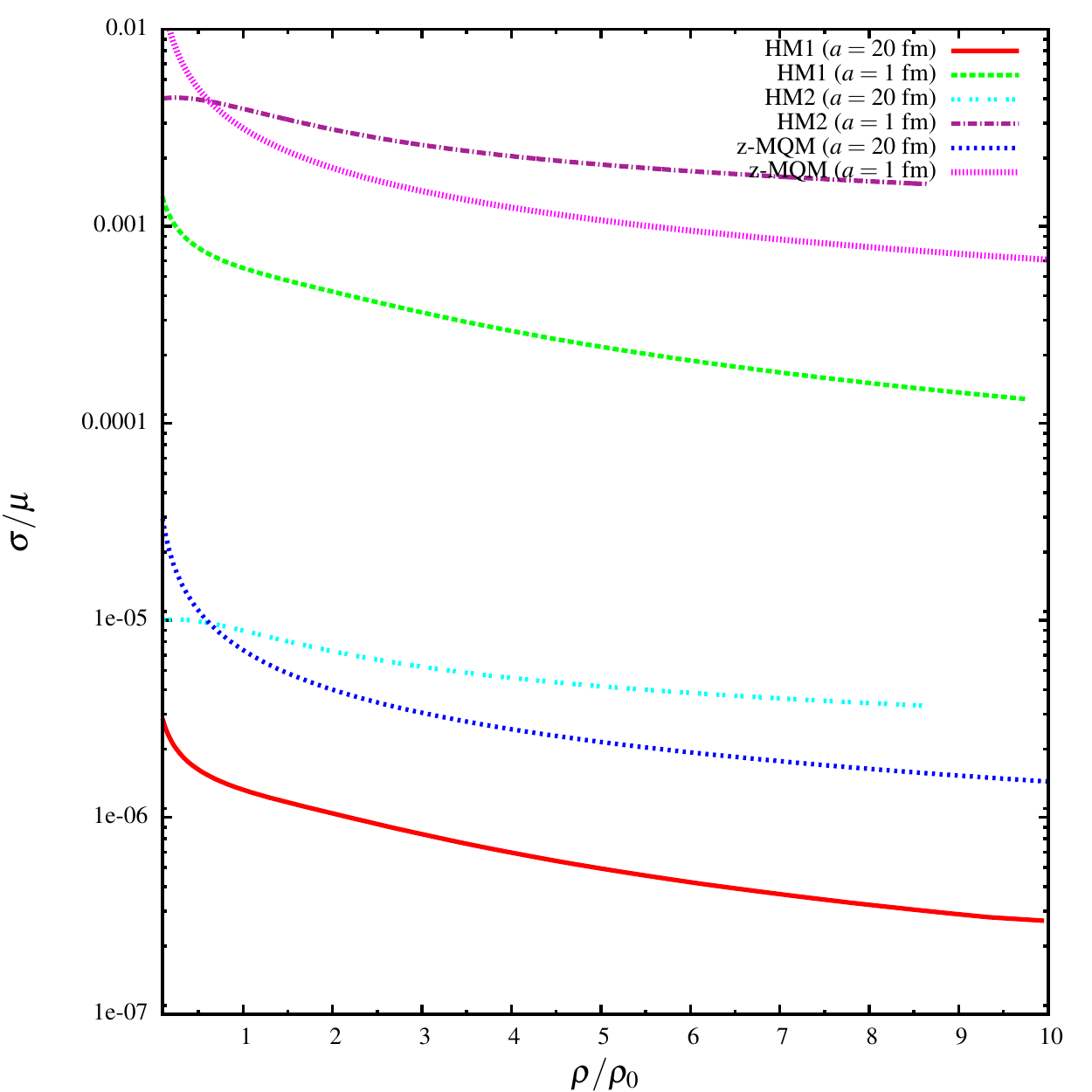}}
\caption{Shear viscosity to density ratio (left) and electrical conductivity upon chemical potential (right) with respect to scaled baryon density of hadronic matter and zero-mass quark matter (z-MQM) for different values of cross-section.}
\label{es_mu_tc}
\end{figure}

 Once we obtain the relaxation time for each phase, we calculate the shear viscosity and the electrical conductivity by using the density dependent relaxation time values. The normalized estimations of $\eta/\rho$ and $\sigma/\mu$ are presented in the left and right panels of Fig.~(\ref{es_mu_tc}), respectively. Though $\eta/\mu^3$ is also possible dimensionless quantity but $\eta/\rho$ is chosen for fluid property~\cite{Gale} measurement of quark and hadronic matter in the dense sector. This ratio $\eta/\rho=\eta\mu/(\epsilon+P)$ at $T= 0$, $\mu\neq 0$ can have equivalent role like viscosity to entropy density ratio $\eta/s$ for early universe or RHIC/LHC environment with $T\neq 0$, $\mu=0$. From the Euler's thermodynamical relation, we get a general relation $Ts=\epsilon+P -\mu\rho$, which will be transformed into $Ts=\epsilon+P$ for $T\neq 0$, $\mu=0$ and $\mu \rho=\epsilon+P$ for $T= 0$, $\mu\neq 0$ respectively. So, for measuring the fluid property, we may choose dimensionless ratio either $\eta T/(\epsilon +P)$ or $\eta \mu/(\epsilon +P)$ for intermediate $T$, $\mu$. Now, when one goes from the intermediate $T$-$\mu$ domain to the limit of $T\neq 0$, $\mu=0$, then $\eta T/(\epsilon +P)$ will be better quantity~\cite{Gale} and in limiting case, we can write $\eta T/(\epsilon +P)=\eta/s$. Similarly, when one goes from the intermediate $T$-$\mu$ domain to the limit of $T=0$, $\mu\neq 0$, then $\eta \mu/(\epsilon+P)$ will be the better quantity to study and in limiting case, we can write $\eta \mu/(\epsilon+P)=\eta/\rho$. Reader may find good discussion on this fluidity quantity in \cite{Gale,Koch_eta}.

 We know that the data of RHIC and LHC experiments~\cite{RHIC1,RHIC2,LHC1,Hydro1,Hydro2} indicate very small values of $\eta/s$, ever observed in nature~\cite{Mclaren}. On the other hand, a string theory-based calculation~\cite{KSS} tells $\frac{\eta}{s}\geq \frac{1}{4\pi}$, which may be considered as lower bound conjecture of $\eta/s$. This lower bound is popularly known as the Kovtun-Son-Starinet (KSS) bound~\cite{KSS}. The existence of a non-zero (but may not be equal to $\frac{1}{4\pi}$) lower bound of $\eta/s$ can also be realized from quantum aspects, which can prevent the classical possibility $\eta/s\rightarrow 0$. Being roughly proportional to the ratio of mean free path to de-Broglie wavelength of medium constituent, the ratio $\eta/s$ of any fluid can never vanish because the mean free path of any constituent can never be lower than its de-Broglie wavelength. It indicates that quantum fluctuations prevent the existence of perfect fluid in nature and $\eta/s$ of any fluid should have some lower bound. Along with the support from quantum aspect, this lower bound of $\eta/s$ is also validated through supersymmetric Yang–Mills (SYM) theory~\cite{SYM}, which provide $\frac{\eta}{s}=\frac{1}{4\pi}$ in the infinite limit of the 't Hooft coupling. From experimental direction also, we noticed that $\eta/s$ of all fluids, including super-fluid helium and even trapped Li atom at strong coupling~\cite{Li}, remain well above the bound for the range of measured temperatures and pressures. All these directions, ultimately support the lower bound conjecture of $\eta/s$, which will be used as a gross reference point in present work. The term "gross" is used because we have to accept the possibility of violation of the KSS bound~\cite{KSS_v1,KSS_v2}.

 Interestingly, the values of $\eta/s$ for RHIC and LHC matter are very close to this quantum lower bound $1/(4\pi)$. So, a natural question arises - whether this nearly perfect fluid nature is also expected along the $\mu$-axis at $T=0$ like that along the $T$-axis at $\mu=0$ of the QCD phase diagram? The present investigation aims in this direction for the first time but to get more conclusive outcome, probably further (alternative) research is required in future. The question becomes more important as recently~\cite{KSS_NM} have experimentally found the value $\eta/s\approx 1/(4\pi)$ for hadronic matter in the low energy hadronic physics experiment. We have plotted $\eta/\rho$ of hadronic and quark matter with two values of scattering length in the left panel of Fig.~(\ref{es_mu_tc}). We notice that the results for $a\approx 20$ fm cross the KSS values ($\approx 0.08$), which may not be considered as a physically acceptable order of magnitude. Probably the magnitude of the cross section/scattering length in vacuum may be largely modified in the finite density picture, which is missing in the present calculation. In this regard, our guess value scattering $a=1$ fm provides an acceptable range of $\eta/\rho$ in the hadronic density range, which still crosses the KSS boundary in the quark density range. So a safe zone may be considered as scattering lengths $a<1$ fm for getting $\eta/\rho>1/(4\pi)$. For mathematical guidance we can get a lower limit curve of the relaxation time $\tau$ as a function of density or chemical potential for massless quark matter by imposing the KSS limit. For finite temperature we get $\tau=5/(4\pi T)$~\cite{JD_Pramana} by imposing KSS limit $\eta/s=1/(4\pi)$. Similarly, for finite density, one can easily find the corresponding expression $\tau=5/(4\pi \mu)$ by using $\eta$ from Eq.~(\ref{es_QM}), $\rho$ from Eq.~(\ref{density_upq}) and then by imposing KSS limit $\eta/\rho=1/(4\pi)$. In terms of density $\rho$, the KSS limit of relaxation time for massless quark matter with $N_f=2$ flavor will be
 
\begin{eqnarray} 
\tau=\frac{5}{4\pi}\Big(\frac{2}{\pi^2}\Big)^{1/3}\frac{1}{\rho^{1/3}}
\end{eqnarray}

Another interesting outcome here is that we still see the jump in $\eta/\rho$ is possible in case of a possible scenario of hadron-quark phase transition as in terms of $\eta/\rho$ the ranking is noticed as HM2 $>$ HM1 $>$ z-MQM. So phase transition from hadronic to quark phase may imply a reduction in $\eta/\rho$. However, for normalized electrical conductivity $\sigma/\mu$, we observed that the ranking is HM2 $>$ z-MQM $>$ HM1. So the ranking of magnitude of $\sigma/\mu$ for the two different phases may be model dependent. We cannot get any conclusive picture in this regard. The present work reveals this uncertainty. We also do not go into any comparative discussion of two hadronic models, rather we intend to show that a possible numerical uncertainty pertains to the estimations of $\sigma/\mu$ by using two different hadronic models. Hence, future research with other existing hadronic models is probably necessary for getting any conclusive or broader picture. Only an order of magnitude difference in transport coefficients for two phases can be considered as conclusive message in present study.

 
\section{Summary and Conclusion}
\label{sec:Sum}

To summarize, we have attempted to visualize the variation of shear viscosity and electrical conductivity of hadronic and quark matter along the density axis. Inspiration of such calculation comes from the calculations of transport coefficients of RHIC or LHC matter, where we get a cross-over type of quark-hadron phase transition along temperature axis. The temperature dependence of shear viscosity and electrical conductivity of RHIC or LHC matter mainly contain two parts - one is thermodynamical phase-space part and the relaxation time part. If we exclude the relaxation time part by normalizing it, then we obtain their nice pattern along temperature axis as one notices for other thermodynamical quantities like pressure, energy density etc. from lattice QCD calculations. The pattern of normalized thermodynamical quantities and transport coefficients is as follows. At high temperature, they reach close to their massless limits and they are gradually suppressed as temperature goes down. Similar kind of pattern is observed in the present work when we go from high density quark phase to low density hadronic phase. The finite density calculations of hadronic phase is obtained with two RMF models and massless quark model for the quark phase. The equivalence of thermodynamic phase-space profile of transportation along temperature axis and density axis is a very interesting finding of the present work. 

 Instead of normalizing relaxation time, if we use its microscopic estimated values, which is generally of the order of fm due to strong interaction, then we can get the final profile of the transport coefficients. A long list of references can be found for RHIC or LHC matter, where most of them found that shear viscosity to entropy density ratio will decrease first then increase with temperature. In this regard, we find shear viscosity to density ratio decreases with density in the hadronic phase and may suddenly drop in a possible scenario of hadron-quark phase transition at $T=0$. In the pure quark phase, a mild decreasing trend with density is observed. This is the qualitative trend of the fluidity measurement of quark and hadronic matter along the density axis, observed in the present investigation. During the computation of the quantitative values, we use the quantum lower bound of fluidity or shear viscosity to density ratio as a physical reference point and we found an interesting finding which is as follows. The experimental data for standard nucleon-nucleon scattering length in vacuum is of the order of 20 fm and using this as a hard-sphere scattering cross section, we find that the fluidity values cross the lower bound. We also found that scattering length less than 1 fm can only provide physical fluidity, beyond its lower bound. It indicates that a good amount of in-medium modification will play an important role in the dense sector, for which vacuum scattering length 20 fm will be reduced to lower values, less than 1 fm. By using scattering length as an input parameter, the present work just unfolds this issue but not well equipped for the microscopic calculation of density dependent scattering length, which may be explored in future. We believe that other hadronic models, whose mass and chemical potential undergo almost similar kind of modification in terms of order of magnitude, will face same problem of fluidity calculation in the dense sector, which should be checked with other hadronic models.


\ack

The authors thank Dr. Sandeep Chatterjee, Department of Physical Sciences, IISER Berhampur for useful discussions. DS acknowledges the institute postdoctoral funding and research facilities at Indian Institute of Science Education and Research Berhampur, Odisha at the initial stage of this work. NA acknowledges the funding support of IFCPAR/CEFIPRA under Project No. 5804-3.



\section*{References}

\begin{thebibliography}{100}

\bibitem{LQCD_sign} P. de Forcrand, {\it{Simulating QCD at finite density}}, PoS LAT 2009 (2009) 010.
%
\bibitem{LQCD_88} F. Karsch, {\it Lattice QCD at finite temperature: a status report}, Zeitschrift Fur Physik C38, 147 (1988).
%
\bibitem{LQCD_2010} S. Borsanyi, G. Endrodi, Z. Fodor et al., {\it The QCD equation of state with dynamical quarks}, Journal of High Energy Physics, 2010, 77 (2010).
%
\bibitem{LQCD_21} J. N. Guenther, {\it Overview of the QCD phase diagram}, Eur. Phys. J. A57, 136 (2021). 
%
\bibitem{HRG} A. N. Tawfik, {\it Equilibrium Statistical-Thermal Models in High-Energy Physics}, Int. J. Mod. Phys. A29 (2014), 1430021.
%
\bibitem{PQCD1} M. Strickland, J. O. Andersen, L. E. Leganger, N. Su, {\it Hard-thermal-loop QCD Ther-modynamics}, Prog. Theor. Phys. Suppl.187, 106 (2011).
%
\bibitem{PQCD2} J. O. Andersen, M. Strickland, and N. Su, {\it Three-loop HTL gluon thermodynamics at intermediate coupling}, JHEP 1008, 113 (2010).
%
\bibitem{NS_QM1} E. Annala, T. Gorda, A. Kurkela, J. Nattila, and A. Vuorinen, {\it Quark-matter cores in neutron stars}, Nature Physics 16, 907 (2020).
%
\bibitem{PQCD_mu1}A. Kurkela, P. Romatschke and A. Vuorinen, {\it{Cold quark matter}}, Phys. Rev. D 81 (2010) 105021.
%
\bibitem{PQCD_mu2}A. Kurkela and A. Vuorinen, {\it{Cool Quark Matter}} Phys. Rev. Lett. 117 (2016) no.4, 042501.
%
\bibitem{PQCD_mu3} T. Gorda, A. Kurkela, P. Romatschke, M. Säppi and A. Vuorinen, {\it Next-to-Next-to-Next-to-Leading Order Pressure of Cold Quark Matter: Leading Logarithm}, Phys. Rev. Lett. 121 (2018) no.20, 202701.
%
\bibitem{Chodos} A. Chodos et al., {\it{New extended model of hadrons}}, Phys. Rev. D 9, 3471 (1974).
%
\bibitem{Sen2} D. Sen and T. K. Jha, {\it{Effects of hadron-quark phase transition on properties of Neutron Stars}}, J. Phys. G: Nucl. Part. Phys. 46 (2019) 015202.
%
\bibitem{Sen} D. Sen and T. K. Jha, {\it{Deconfinement of nonstrange hadronic matter with nucleons and $\Delta$ baryons to quark matter in neutron stars}}, Int. J. Mod. Phys. D 28 (2019) no.02, 1950040.
%
\bibitem{Sen3} D. Sen et al., {\it{Properties of Neutron Stars with hyperon cores in parameterized hydrostatic conditions}}, Int.J.Mod.Phys. E27 (2018) 1850097.
%
\bibitem{Sen5} D. Sen, {\it{Nuclear matter at finite temperature and static properties of proto-neutron star}}, J.Phys. G48 (2021) 025201.
%
\bibitem{Sen6} D. Sen, {\it{Variation of the $\Delta$ baryon mass and hybrid star properties in static and rotating conditions}}, Phys.Rev.C 103 (2021) 4, 045804; 	arXiv:2103.14136 [nucl-th].
%
\bibitem{Sen4} D. Sen, {\it{Role of $\Delta$s in determining the properties of neutron stars in parameterized hydrostatic equilibrium}}, Int. J. Mod. Phys. D 28, No. 9 (2019) 1950122.
%
\bibitem{Sahu2000}P. K. Sahu and A. Ohnishi, {\it{SU(2) Chiral Sigma Model and Properties of Neutron Stars}}, Prog. Theor. Phys. 104, 1163 (2000).
%
\bibitem{Sahu2004}P. K. Sahu, T. K. Jha, K. C. Panda, and S. K. Patra, {\it{Hot Nuclear Matter in Asymmetry Chiral Sigma Model}}, Nucl. Phys. A733, 169 (2004).
%
\bibitem{TKJ}T. K. Jha and H. Mishra, {\it{Constraints on nuclear matter parameters of an effective chiral model}}, Phys. Rev. C 78 (2008) 065802.
%
\bibitem{Agrawal12}B. K. Agrawal, A. Sulaksono and P. -G. Reinhard, {\it{Optimization of relativistic mean field model for finite nuclei to neutron star matter}}, Nucl. Phys. A882, 1-20 (2012).
%
\bibitem{Sulaksono14} A. Sulaksono, N. Alam and B. K. Agrawal, {\it{Core–crust transition properties of neutron stars within systematically varied extended relativistic mean-field model}}, Int. J. Mod. Phys. E 23, 1450072 (2014).
%
\bibitem{Alam15} N. Alam, A. Sulaksono, and B. K. Agrawal, {\it{Diversity of neutron star properties at the fixed neutron-skin thickness of $^{208}\mathrm{Pb}$}}, Phys. Rev. C 92, 015804 (2015).
%
\bibitem{Alam17}N. Alam, H. Pais, C. Provid\^encia, and B. K. Agrawal, {\it{Warm unstable asymmetric nuclear matter: Critical properties and the density dependence of the symmetry energy}}, Phys. Rev. C 95, 055808 (2017).
%
\bibitem{CAI_JD_SG}C. A. Islam, J. Dey, S. Ghosh, {\it Impact of different extended components of mean field models on transport coefficients of quark matter and their causal aspects}, Phys. Rev. C103 (2021) 3, 034904
%
\bibitem{1950} L. Mestel and F. Hoyle, {\it On the thermal conductivity in dense stars}, Proc. Cambridge Philos. Soc. 46, 331 (1950).
%
\bibitem{1964}A. A. Abrikosov, {\it The conductivity of strongly compressed matter}, Sov. Phys. JETP 18, 1399 (1964).
%
\bibitem{1970}V. Canuto, {\it Electrical conductivity and conductive opacity of a relativistic electron gas}, Astrophys. J. 159, 641 (1970).
%
\bibitem{1976}E. Flowers and N. Itoh, {\it Transport properties of dense matter},
Astrophys. J. 206, 218 (1976).
%
\bibitem{Baym1969} G. Baym, C. Pethick, and D. Pikes, {\it{Electrical Conductivity of Neutron Star Matter}}, Nature volume 224, pages674–675 (1969).
%
\bibitem{Yakovlev} D. G. Yakovlev and D. A. Shalybkov, {\it{Electrical conductivity of neutron star cores in the presence of a magnetic field}}, Astrophysics and Space Science volume 176, pages191–215 (1991).
%
\bibitem{Ewart} G. M. Ewart, R. A. Guyer, and G. Greenstein, {\it{Electrical conductivity and magnetic field decay in neutron stars}}, The Astrophysical Journal, 202:238-247, 1975.
%
\bibitem{Raikh} M. E. Raikh and D. G. Yakovlev, {\it{Thermal and electrical conductivities of crystals in neutron stars and degenerate dwarfs}}, Astrophysics and Space Science volume 87, pages193–203 (1982)
%
\bibitem{Potekhin1999} A. Y. Potekhin et al., {\it{Transport properties of degenerate electrons in neutron star envelopes and white dwarf cores}}, Astron.Astrophys.346:345,1999.
%
\bibitem{Shternin2008} P. S. Shternin and D. G. Yakovlev, {\it{Shear viscosity in neutron star cores}},
Phys. Rev. D 78, 063006 (2008)
%
\bibitem{Banik2010} S. Banik and D. Bandyopadhyay, {\it{Effect of shear viscosity on the nucleation of antikaon condensed matter in neutron stars}}, Phys.Rev.D82:123010,2010
%
\bibitem{Banik2011} S. Banik, R. Nandi, and D. Bandyopadhyay, {\it{Shear viscosity and the nucleation of antikaon condensed matter in protoneutron stars}} Phys.Rev.C 84 (2011) 065804.
%
\bibitem{Schmitt2017} A. Schmitt and P. Shternin, {\it{Reaction Rates and Transport in Neutron Stars}}, Astrophys.Space Sci.Libr. 457 (2018) 455-574.
%
\bibitem{Shternin2020} P. Shternin and M. Baldo, {\it{Transport coefficients of nucleon neutron star cores for various nuclear interactions within the Brueckner-Hartree-Fock approach}}, Phys. Rev. D 102, 063010 (2020).
%
\bibitem{McLaughlin} E. McLaughlin et al., {\it{Building a testable shear viscosity across the QCD phase diagram}}, arXiv:2103.02090 [nucl-th] (2021).
%
\bibitem{Shternin2013} P. Shternin, M. Baldo, and P. Haensel, {\it{Transport coefficients of nuclear matter in neutron star cores}}, Phys.Rev.C 88 (2013) 6, 065803.
%
\bibitem{Shternin2017} P. Shternin, M. Baldo, and H-J Schulze, {\it{Transport coefficients in neutron star cores in BHF approach. Comparison of different nucleon potentials}}, J. Phys.: Conf. Ser. 932, 012042 (2017).
%
\bibitem{Shang2020} X. L. Shang, P. Wang, W. Zuo, and J. M. Dong, {\it{Role of nucleon-nucleon correlation in transport coefficients and gravitational-wave-driven r-mode instability of neutron stars}}, Phys. Lett. B, 811, 135963 (2020).
%
\bibitem{Nandi2018} R. Nandi, and S. Schramm, {\it{Calculation of the transport coefficients of the nuclear pasta phase}}, J.Astrophys.Astron. 39 (2018) 40.
%
\bibitem{Shternin2021} P. Shternin and I. Vidana, {\it{Transport Coefficients of Hyperonic Neutron Star Cores}}, Universe 7(6), 203 (2021).
%
\bibitem{Shternin_18}P. S. Shternin, Phys. Rev. D 98, 063015 (2018)

\bibitem{VolKo}V. Koch, {\it{Aspects of Chiral Symmetry}}, International Journal of Modern Physics E6, 203-250 (1997).
%
\bibitem{Sahu1993}P. K. Sahu, R. Basu, and B. Datta, {\it{High-Density Matter in the Chiral Sigma Model}} Astrophys. J. 416, 267 (1993).
%
\bibitem{Stone2} J. R. Stone and P. G. Reinhard, {\it {The Skyrme Interaction in finite nuclei and nuclear matter}} Prog. Part. Nucl. Phys. 58, 587 (2007).
%
\bibitem{Khan2013} E. Khan and J. Margueron, {\it{Determination of the density dependence of the nuclear incompressibility}}, Phys. Rev. C 88, 034319 (2013).
%
\bibitem{Garg} U. Garg and G. Colo, {\it{The compression-mode giant resonances and nuclear incompressibility}}, Prog. Part. Nucl. Phys. 101 (2018) 55.
%
\bibitem{Dutra2014} M. Dutra et. al., {\it{Relativistic mean-field hadronic models under nuclear matter constraints}}, Phys. Rev., C 90, 055203 (2014).
%
\bibitem{Stone} J. R. Stone, N. J. Stone, S. A. Moszkowski, {\it{Incompressibility in finite nuclei and nuclear matter}}, Phys. Rev. C 89, 044316 (2014).
%
\bibitem{Fattoyev} F. J. Fattoyev et al., {\it{Neutron Skins and Neutron Stars in the Multimessenger Era}}, Phys. Rev. Lett. 120, 172702 (2018).
%
\bibitem{Zhu2018} Z-Y Zhu, E-P Zhou, A. Li {\it{Neutron star equation of state from the quark level in the light of GW170817}}, Astrophys.J. 862 (2018) no.2, 98.
%
\bibitem{Sen7} D. Sen, N. Alam, and G. Chaudhuri, {\it{Properties of hybrid stars with a density-dependent bag model}}, J. Phys. G: Nucl. Part. Phys. 48 (2021) 105201.
%
\bibitem{Walecka74} J. D. Walecka, {\it{A theory of highly condensed matter}}, Annals of Physics, 83, 491 (1974).
%
\bibitem{Boguta77} J. Boguta and A. R. Bodmer, {\it{Relativistic calculation of nuclear matter and the nuclear surface}}, Nucl. Phys. A292, 413 (1977).
%
\bibitem{Serot86} B. D. Serot and J. D. Walecka, {\it{Advances in Nuclear Physics}}, edited by J. W. Negele and E. Vogt (Plenum, New York, 1986), Vol. 16.
%
\bibitem{Furnstahl87} R. J. Furnstahl, C. E. Price, and G. E. Walker, {\it{Systematics of light deformed nuclei in relativistic mean-field models}}, Phys. Rev. C 36, 2590 (1987).
%
\bibitem{Muller96} H. Muller and B. D. Serot, {\it{Relativistic mean-field theory and the high-density nuclear equation of state}}, Nucl. Phys. A 606, 508 (1996)
%
\bibitem{Lalazissis97} G. A. Lalazissis, J. K\"onig, and P. Ring, {\it{New parametrization for the Lagrangian density of relativistic mean field theory}}, Phys. Rev. C 55, 540 (1997).
%
\bibitem{Serot97} B. D. Serot and J. D. Walecka, {\it{Recent Progress in Quantum Hadrodynamics}}, Int. J. Mod. Phys. E 6, 515 (1997).%
%
\bibitem{Glendenning} N. K. Glendenning, {\it Compact Stars: Nuclear Physics, Particle Physics, and General Relativity} (Springer-Verlag, New York, 2000).
%
\bibitem{Kapusta}P. Chakraborty, J.I. Kapusta, 
{\it Quasi-Particle Theory of Shear and Bulk Viscosities of Hadronic Matter}
Phys.Rev.C 83 (2011) 014906
%
\bibitem{Gavin} S. Gavin, {\it Transport Coefficients in ultra-relativistic heavy ion collision} Nucl. Phys. A 435 (1985) 826. 
%
\bibitem{HTL_3L} M. Strickland, J. O. Andersen, L. E. Leganger, N. Su, {\it Hard-thermal-loop QCD Ther-modynamics},
Prog. Theor. Phys. Suppl.187, 106 (2011).
%
\bibitem{Hatsuda}G. Baym, T. Hatsuda, T. Kojo, P. D. Powell, Y. Song, T. Takatsuka, {\it From hadrons to quarks in neutron stars: a review}, Reports on Progress in Physics 81 (2018) 056902.
%
\bibitem{NN1} M. M. Nagelset al., {\it{Compilation of coupling constants and low-energy parameters}} Nucl. Phys. B147, 189 (1979).
%
\bibitem{NN2} O. Dumbrajset al., {\it{Compilation of coupling constants and low-energy parameters. 1982-edition}}, Nucl. Phys. B216, 277 (1983).
%
\bibitem{SB_SG2} S. Ghosh, S. Ghosh, S. Bhattacharya, {\it{Phenomenological bound on the viscosity of the hadron resonance gas}}, Phys. Rev. C98 (2018) 045202.
%
\bibitem{Gale} G.S. Denicol, C. Gale, S. Jeon, J. Noronha, {\it Fluid behavior of a baryon-rich hadron resonance gas}, Phys. Rev. C 88 (2013) 6, 064901
%
\bibitem{RHIC1} PHENIX collaboration, S. S. Adler et al., 
{\it Elliptic flow of identified hadrons in Au+Au collisions at $\sqrt{s_{NN}}= 200$ GeV}, Phys. Rev. Lett. 91 (2003) 182301,
%
\bibitem{RHIC2} STAR collaboration, J. Adams et al., 
{\it Azimuthal anisotropy in Au+Au collisions at $\sqrt{s_{NN}}= 200$ GeV}, Phys. Rev. C 72 (2005) 014904
%
\bibitem{LHC1} ALICE collaboration, K. Aamodt et al., 
{\it Higher harmonic anisotropic flow measurements of charged particles in Pb-Pb collisions at $\sqrt{s_{NN}}= 2.76$ TeV}, 
Phys. Rev. Lett. 107 (2011) 032301.
%
\bibitem{Hydro1}P. Romatschke and U. Romatschke, 
{\it Viscosity Information from Relativistic Nuclear Collisions: How Perfect is the Fluid Observed at RHIC?}, Phys. Rev. Lett. 99, 172301 (2007)
%
\bibitem{Hydro2}U. Heinz and R. Snellings, {\it Collective flow and viscosity in relativistic heavy-ion collisions}, Annu. Rev. Nucl. Part. Sci. 63 (2013) 123.
%
\bibitem{Mclaren}M. Gyulassy and L. McLerran, Nucl. Phys. A 750, 30 (2005).
%
\bibitem{KSS}P. Kovtun, D. T. Son, and O. A. Starinets, 
Phys. Rev. Lett. 94, 111601 (2005).
%
\bibitem{KSS_NM}D. Mondal et al. Phys. Rev. Lett. 118, 192501 (2017).
%
\bibitem{JD_Pramana}J. Dey, S. Satapathy, P. Murmu, S. Ghosh {\it Shear viscosity and electrical conductivity of relativistic fluid in presence of magnetic field: a massless case}, Pramana 95 (2021) 3, 125

\bibitem{SYM} A. Buchel, J. T. Liu and A. O. Starinets, {\it Coupling constant dependence of the shear viscosity in N=4 supersymmetric Yang-Mills theory}, Nucl. Phys. B 707, 56 (2005) 
%
\bibitem{Li}T. Schafer, Phys. {\it Ratio of shear viscosity to entropy density for trapped fermions in the unitarity limit}, Rev. A 76, 063618 (2007).
%
\bibitem{KSS_v1}A. Dobado and F. J. Llanes-Estrada, {\it On the violation of the holographic viscosity versus entropy KSS bound in non relativistic systems}, Eur. Phys. J. C 51, 913 (2007).
%
\bibitem{KSS_v2}R. C. Myers, M. F. Paulos and A. Sinha, {\it Holographic Hydrodynamics with a Chemical Potential}, JHEP 0906, 006 (2009).
%
\bibitem{Koch_eta}J. Liao, V. Koch, {\it Fluidity and supercriticality of the QCD matter created in relativistic
heavy ion collisions}, Phys. Rev. C 81, 014902 (2010).


\end{thebibliography}
\end{document}